\documentclass[twocolumn]{aastex63}
\usepackage{paperstyle}

\AtEndEnvironment{threeparttable}{\vskip10pt}{}{}

\definecolor{darkgreen}{rgb}{0.0, 0.55, 0.0}
\definecolor{purple}{rgb}{0.75, 0.0, 1.0}
\def\msun{\hbox{${ M}_{\odot}$}}
\def\mstar{\hbox{${M}_{\star}$}}

\received{October 17, 2021}
\revised{April 25, 2022}
\accepted{May 15, 2022}

\submitjournal{ApJ}

\shorttitle{Stellar mass profiles of galaxies in the Hubble Frontier Fields}
\shortauthors{Tan et al.}

\begin{document}

\author[0000-0002-3503-8899]{Vivian Yun Yan Tan}
\affiliation{Department of Physics and Astronomy, York University, 4700 Keele Street, Toronto, ON, M3J 1P3, Canada}
\author[0000-0002-9330-9108]{Adam Muzzin}
\affiliation{Department of Physics and Astronomy, York University, 4700 Keele Street, Toronto, ON, M3J 1P3, Canada}
\author[0000-0002-7248-1566]{Z. Cemile Marsan}
\affiliation{Department of Physics and Astronomy, York University, 4700 Keele Street, Toronto, ON, M3J 1P3, Canada}
\author[0000-0003-0780-9526]{Visal Sok}
\affiliation{Department of Physics and Astronomy, York University, 4700 Keele Street, Toronto, ON, M3J 1P3, Canada}
\author[0000-0002-2250-8687]{Leo Y. Alcorn}
\affiliation{David A. Dunlap Department of Astronomy \& Astrophysics, University of Toronto, 50 St. George St., Toronto, ON M5S 3H4, Canada}
\author[0000-0002-7547-3385]{Jasleen Matharu}
\affiliation{Department of Physics and Astronomy, Texas A\&M University, College Station, TX, 77843-4242, USA\\}
\affiliation{George P. and Cynthia Woods Mitchell Institute for
 Fundamental Physics and Astronomy, Texas A\&M University, College Station, TX, 77845-4242, USA\\}
\author{Heath Shipley}
\affiliation{Department of Physics, McGill University, 3600 Rue University, Montr\'eal, QC H3P 1T3, Canada}
\author[0000-0001-9002-3502]{Danilo Marchesini}
\affiliation{Department of Physics and Astronomy, Tufts University, 574 Boston Avenue, MA 02155, USA}
\author[0000-0001-5294-8002]{Kalina~V.~Nedkova} 
\affiliation{Department of Physics and Astronomy, Tufts University, 574 Boston Avenue, MA 02155, USA}
\author[0000-0003-3243-9969]{Nicholas Martis}
\affiliation{Astronomy and Physics Department, St. Mary’s University, Halifax, NS}
\affiliation{National Research Council of Canada, Herzberg Astronomy \& Astrophysics Research Centre, 5071 West Saanich Road, Victoria, BC, Canada, V9E2E7}
\author[0000-0002-5027-0135]{Arjen van der Wel}
\affiliation{Dept. of Physics \& Astronomy, Ghent University, Krijgslaan 281, Building S9, 9000 Ghent, Belgium}
\author[0000-0001-7160-3632]{Katherine E. Whitaker}
\affiliation{Department of Astronomy, University of Massachusetts, 710 North Pleasant St, LGRT-524, Amherst, MA 01003, USA}

\renewcommand{\thefootnote}{\fnsymbol{footnote}}
\title{Resolved Stellar Mass Maps of Galaxies in the Hubble Frontier Fields:  Evidence for Mass Dependency in Environmental Quenching}

\begin{abstract}
 One of the challenges in understanding the quenching processes for galaxies is connecting progenitor star-forming populations to their descendant quiescent populations over cosmic time. Here we attempt a novel approach to this challenge by assuming that the underlying stellar mass distribution of galaxies is not significantly altered during environmental quenching processes that solely affect the gas content of cluster galaxies, such as strangulation and ram-pressure stripping. Using the deep, high-resolution photometry of the Hubble Frontier Fields, we create resolved stellar mass maps for both cluster and field galaxies, from which we determine 2D S\'ersic profiles, and obtain S\'ersic indices and half-mass radii. We classify the quiescent cluster galaxies into disk-like and bulge-like populations based on their S\'ersic indices, and find that bulge-like quiescent galaxies dominate the quiescent population at higher masses ($\mstar > 10^{9.5}\msun$), whereas disk-like quiescent galaxies dominate at lower masses ($10^{8.5}\msun< \mstar < 10^{9.5}\msun$).  Using both the S\'ersic indices and half-mass radii, we identify a population of quiescent galaxies in clusters that are ``morphological analogues" of field star-forming galaxies. These analogues are interpreted to be star-forming galaxies that had been environmentally quenched.  We use these morphological analogues to compute the environmental-quenching efficiency, and we find that the efficiency decreases with increasing stellar mass. This demonstrates that environmental quenching is more effective on less massive galaxies and that the effect of environment on quenching galaxies is not completely separable from the effect of mass on quenching galaxies.
\end{abstract}

\section{Introduction}\label{chap:intro}

Galaxy clusters are the largest collapsed structures in the universe, and the high density of galaxies inside them means they are host to a diversity of physical processes that affect galaxy evolution. These physical processes result in clusters containing higher numbers of red-sequence or  ``quenched" galaxies, which are galaxies that typically have a red colour and low star-formation rates \citep{K2004, B2004a,B2006,S2006,V2008,P2010}. Indeed, environmental density and its connection to quenching is one of the most well-studied relationships in galaxy evolution over the last several decades (\citealt{GG1972, D1980, PG1984, BV1990}, see \citealt{BG2006,BM2009} for a review). 

In a key paper, \cite{P2010} proposed that there are two channels to quenching galaxies, one related to their stellar mass (``mass quenching") and one related to their environment (``environmental quenching").
\cite{P2010} analyzed data from the Sloan Digital Sky Survey (SDSS) and the two main results of their formalism showed that: 1) mass quenching operates at the same efficiency across all environmental densities, and 2) environmental quenching only increases in proportion to the environmental number density, and is independent of stellar mass. \cite{P2010} also uses the zCOSMOS survey to show their model holds at redshifts up to $z\sim 0.7$, and other early studies confirmed their results to $z\sim 1$ (\citealt{M2012, Q2012, K2014}).  Interestingly, more recent studies that focus on redshifts in the range of $z \sim0.5 $ to $z\sim 1$ and beyond, find increasing evidence that environmental quenching {\it does not} solely depend on environmental density, but also on the stellar mass of the galaxy \citep{B2016,D2016,K2017,Li2017,P2019, vanderBurg:2020,Cutler:2022}. These somewhat contradictory results raise difficult questions, and determining whether a galaxy has quenched because of internal processes, or because of the influences of its environment remains one of the most important outstanding questions in galaxy evolution.

Thus far there have been many empirical measurements of quenched fractions from observations as a function of environment, stellar mass, and redshift. However, there is much less work on directly connecting the {\it physical processes} of quenching to the empirical measurements of quenching relations.  In terms of physical processes, clusters contain hot diffuse gas called the intracluster medium (ICM), which can strip cold gas from galaxies when they fall into the gravitational potential of the cluster, in a process known as ram pressure stripping \citep{GG1972}.  Quenching can also happen due to interactions with other galaxies, such as tidal interactions \citep{M1984, M1996} or starvation/strangulation \citep{L1980}. 

Ram pressure stripping, starvation/strangulation, and tidal stripping are all interactions that mostly affect the gas content of a galaxy and not its stellar content, the latter of which makes up over 50\% of a galaxy’s total baryonic mass at low redshift \citep{Geach:2011,Carilli:2013}. Given that dark matter and stars dominate the gravitational potential of galaxies, this means environmental quenching processes should leave the stellar mass distribution of a galaxy relatively unaffected while altering its gas content and star-formation rate. This hypothesis is borne out in recent hydrodynamical simulations such as by \cite{Ayromlou:2019}, which model both gas and stellar mass in ram-pressure stripping events.  It is also seen in recent resolved studies of galaxies being ram-pressure stripped (e.g. \citealt{Koopmann:2004, Fumagalli:2014, Poggianti:2017a, Fossati:2018, Cortese:2019, Cramer:2019}), where the underlying mass distribution of the galaxy remains largely normal even though spectacular stripping of the gas is observed. 

Why then is there so much difficulty connecting quenched fractions to physical quenching mechanisms?  One of the major issues in all studies of galaxy evolution is the connection of progenitors and descendants, and this issue is no less difficult for cluster studies.  However, if environmental quenching does not alter the underlying stellar mass distribution of galaxies at low redshifts, then we should expect the environmentally-quenched descendants of star-forming galaxies to have the same stellar morphologies as their progenitors. This may also be connected to the observation that red-sequence galaxies have a diversity in morphologies (see review by \citealt{BM2009}), which could be due to the diversity of quenching mechanisms.  Here we posit, that based on the above arguments, the morphology of a quiescent galaxy can be used to trace its progenitors and quenching history if properly measured.

While studying the morphology of galaxies is usually done using integrated light profiles, usually in redder filters, this may not be the ideal representation of the underlying mass distributions. \cite{S2013}, who compared the sizes of light and mass profiles of galaxies, found that the average half-mass radii were ${\sim}15\%$ smaller than their half-light radii. \cite{W2012} and \cite{S2019} uses mass-to-light ratios to study sizes of high redshift galaxies, also noting a difference between the half-mass and half-light sizes.  Indeed, not only does the half-mass radius of galaxies typically differ from their half-light radius, the total stellar mass can be incorrectly estimated if not measured using resolved data as well. For example, \cite{SS2015} fit SEDs to individual pixels of SDSS galaxies in order to compare resolved stellar masses to unresolved stellar mass obtained and found that the stellar masses of star-forming galaxies can be underestimated by up to 25\%.

In this paper we describe a novel method to connect environmentally-quenched quiescent cluster galaxies (descendants) to star-forming field galaxies (progenitors) by quantifying their morphology with resolved stellar mass maps.  We do this by making the assumption that the stellar mass morphology of a galaxy is conserved when a galaxy quenches inside a cluster from environmental processes. We posit that the majority of physical processes involved in environmental quenching should only affect the gas, and not the underlying distribution of the stellar mass throughout the galaxy. Therefore quiescent galaxies with the same total stellar mass and 2D stellar mass profile as field star-forming galaxies are likely to be their environmentally-quenched descendants.  In order to perform this analysis we need a dataset that is multi-wavelength, high spatial resolution, and deep enough to model galaxy SEDs on the pixel level.

The Hubble Frontier Fields \citep{L2017} is ideal for our analysis as it meets all the above criteria and contains observations of six massive clusters at 0.3 $< z <$ 0.6, as well as six complementary flanking fields.  We use the PSF-matched images from the Deep Space collaboration \citep{S2018} as well as their catalogs that contain deep multiband photometry of both cluster and field environments.  The combination of the two allows us to measure both the resolved and unresolved SED of each galaxy as well as the morphology in stellar mass, not just in a single filter.

This paper is structured as follows: the details of the Frontier Fields Deep Space dataset are described in \S \ref{chap:data}, and the process by which the resolved stellar mass distributions are constructed is outlined in \S \ref{chap:method}. The analysis of the mass distributions' Sersic parameters in relation to star-formation, stellar mass, and morphology is in \S \ref{chap:results}, and discussion of the implications of mass dependence in environmental quenching is presented in \S \ref{chap:disc}. Throughout this paper, we adopt a $\Lambda$CDM cosmology with $H_0 = 70$km s$^{-1}$, $\Omega_m = 0.3$, and $\Omega_\Lambda = 0.7$.

\section{Data Sample\label{chap:data}}

\subsection{HFF DeepSpace catalog\label{sec:data1}}

We utilized the Hubble Frontier Fields Deep Space catalog (HFF, release paper \citealt{L2017,S2018}) for this analysis. The galaxy catalog contains six lensing clusters at redshifts of $0.3 < z < 0.6$ and six flanking fields imaged with UV, optical, and NIR with the Advanced Camera for Surveys (ACS) and Wide Field Camera 3 (WFC3) on the Hubble Space Telescope (HST). For this study, we create stellar mass maps using the deep, high-resolution HST imaging. 

The native pixel resolution for WFC3/UVIS is 0.04"~pixel$^{-1}$, for ACS is 0.049"~pixel$^{-1}$, and for WFC3/IR is 0.128"~pixel$^{-1}$. The reddest filter used for HST imaging is F160W on WFC3, which has a PSF FWHM of $\sim0.177"$. The FWHM of the PSF for F814W, the reddest filter on HST's ACS is $\sim 0.099"$. The angular diameter distance at the redshift of the clusters (0.25 $< z <$ 0.6) implies a physical scale of 5.41 -- 9.24 kpc arcsec$^{-1}$.  Based on the WFC3 and ACS PSF size, this implies an angular resolution of 0.96 -- 1.64 kpc per PSF FWHM. 

In addition to the HST data, imaging and photometry is available from both VLT/HAWK-I and Keck/MOSFIRE in the K-band, and Spitzer/IRAC in the 3.6 -- 4.5 $\mu m$ bands.   The average PSF FWHM for the Frontier Fields in the K-band is $\sim 0.4$" \citep{Br2016} and the IRAC PSF FWHM is $\sim1-2$".  These data provide angular resolutions 3 - 10 times worse than the HST imaging and therefore were excluded from the stellar mass map construction.  We note that the derived values of the integrated stellar masses in the Deep Space catalog include this poorer angular resolution photometry.  As we show in $\S$\ref{chap:method}, our total stellar masses derived from just the HST imaging are consistent with the stellar masses determined including the K-band and IRAC data in the DeepSpace catalogs, and therefore the exclusion of those bands does not affect our results.

Although the HFF catalogs contain up to 17 filters across all three cameras, not every cluster was imaged with all the available filters. The number of available filters ranges from 9 for Abell 2744 to 17 for MACS0416 and MACS1149. For the flanking fields, each of the pointings have 7 filters except for the parallel field of MACS0416, which was imaged with 11 filters. Each cluster and parallel field catalog has roughly 5000-8000 objects. Table \ref{data:catalogtable} lists the filters used to image each pointing, and the exact number of objects. We refer the reader to \cite{S2018} (shortened to S18) for complete details on the photometric catalog construction process. 
 
\fontsize{8}{8}\selectfont
\begin{table}
	\caption{\textit{HST bands and sample sizes for each catalog}.\label{data:catalogtable}}
	\begin{tabular*}{\columnwidth}{l l }
	\\
	\hline
	\textbf{ Catalog }& \textbf{Filters Used} \\
	\hline
	Abell370-Cluster & F275W, F336W, F435W, F475W, \\
	Filters: 12 &F606W, F625W, F814W,F105W, \\
	Total objects: 6795 & F110W, F125W,F140W, F160W  \\
	Objects used: 78& \\
	\hline
	
	Abell1063-Cluster & F225W, F275W, F336W, F390W,  \\
	Filters: 16 & F435W, F475W, F606W, F625W,\\
	Total objects:7611 & F775W,  F814W, F850LP, F105W,\\
	Objects used: 27 & F110W, F125W, F140W, F160W \\
	\hline
	
	Abell2744-Cluster & F275W, F336W F435W, \\
	Filters: 9 & F606W, F814W,  F105W, \\
	Total objects: 9390 & F125W, F140W, F160W\\
	Objects used: 127 & \\
	\hline
	MACS0416-Cluster & F225W,F275W, F336W, F390W \\
	Filters: 16 &  F435W,F475W,  F606W, F625W, \\
	Total objects: 7431 &F775W, F814W, F850LP, F105W, \\
	Objects used: 45& F110W, F125W, F140W, F160W \\
	
	\hline
	
	MACS0717-Cluster & F225W, F275W, F336W, F390W,  \\
	Filters: 17 & F435W, F475W, F555W, F606W, \\
	Total objects: 6370 &  F625W, F775W,  F814W, F105W, \\
	Objects used: 42 &  F850LP, F110W, F125W, F140W, \\
	& F160W \\
	\hline
	
	MACS1149-Cluster &  F225W, F275W, F336W, F390W,  \\
	Filters: 17 & F435W, F475W, F555W, F606W,\\
	Total objects: 6868 & F625W, F775W,  F814W, F850LP,\\
	Objects used: 81 & F105W, F110W, F125W, F140W,\\
	& F160W \\
	\hline
	Abell370-Parallel & F435W, F606W, F814W \\
	Filters: 7 & F105W, F125W, F140W, F160W \\
	Total objects: 5687 & \\
	Objects used: 27 & \\
	\hline
	Abell1063-Parallel & F435W, F606W, F814W  \\
	Filters: 7 & F105W, F125W, F140W, F160W   \\
	Total objects: 5574 & \\
	Objects used: 7 & \\
	\hline
	Abell2744-Parallel & F435W, F606W, F814W  \\
	Filters: 7 & F105W, F125W, F140W, F160W  \\
	Total objects: 6240 & \\
	Objects used: 34 & \\
	\hline
	MACS0416-Parallel & F435W, F475W, F606W,  F625W, \\
	Filters: 11& F775W, F814W, F850LP, \\
	Total objects: 7771 & F105W, F125W, F140W, F160W  \\
	Objects used: 14 & \\
	\hline
	MACS0717-Parallel & F435W, F606W, F814W \\
	Filters: 7 & F105W, F125W, F140W, F160W \\
	Total objects: 5776 & \\
	Objects used: 6 & \\
	\hline
	MACS1149-Parallel & F435W, F606W, F814W \\
	Filters: 7 & F105W, F125W, F140W, F160W \\
	Total objects: 5802 & \\
	Objects used: 8 & \\
	\hline
	\end{tabular*}
\end{table}
\normalsize 

\subsection{Sample Selection}\label{sec:data2}

The HFF Deep Space catalog has a 90\% completeness at an AB magnitude for the F160W filter ranging from 28.1 to 28.7 for the deepest region (denoted ``reg1"), and 27.0 to 26.1 for the shallowest region (denoted ``reg3", see S18 for details on how completeness was calculated). In order to determine the limiting stellar mass for our sample, we compare the AB magnitude versus integrated stellar mass. For objects that have a magnitude of 26 in the F160W band, the average integrated stellar mass in the catalogs is $\sim 10^{8.5} \msun$ at our redshift limit of $z\sim 0.6$. In order to retain a highly complete sample, we chose a stellar mass limit of $10^{8.5} \msun$ for the entire sample. This low mass limit allows us to study the environmental effects of low-mass galaxies in the cluster. 

Some of the brightest galaxies in the cluster have significant extended light due to the depth of the imaging. This extended light interferes with photometry of fainter galaxies, especially in the NIR. These galaxies have been modeled out by S18 and are referred to as ``bCGs" or bright cluster galaxies. This means objects that are not the brightest cluster galaxy (BCG, with $\mstar \geq 10^{12} \msun$), such as galaxies with masses of $10^{10.5} \msun - 10^{11.5} \msun$, are considered ``bCGs".  The bCGs were modeled out based on the method outlined in \cite{F2006} (See \S3.1 of \cite{S2018} for more details on bCG modelling for the Frontier Fields' clusters). 
Fortunately the bCG photometry is preserved in the original images and can still be used for analysis after PSF-matching to F160W resolution. Including the ``bCGs" allows us to model a mass-complete population from $10^{8.5} \msun - 10^{12.0} \msun$

\begin{figure*}
	\centerline{\includegraphics[width=0.8\textwidth]{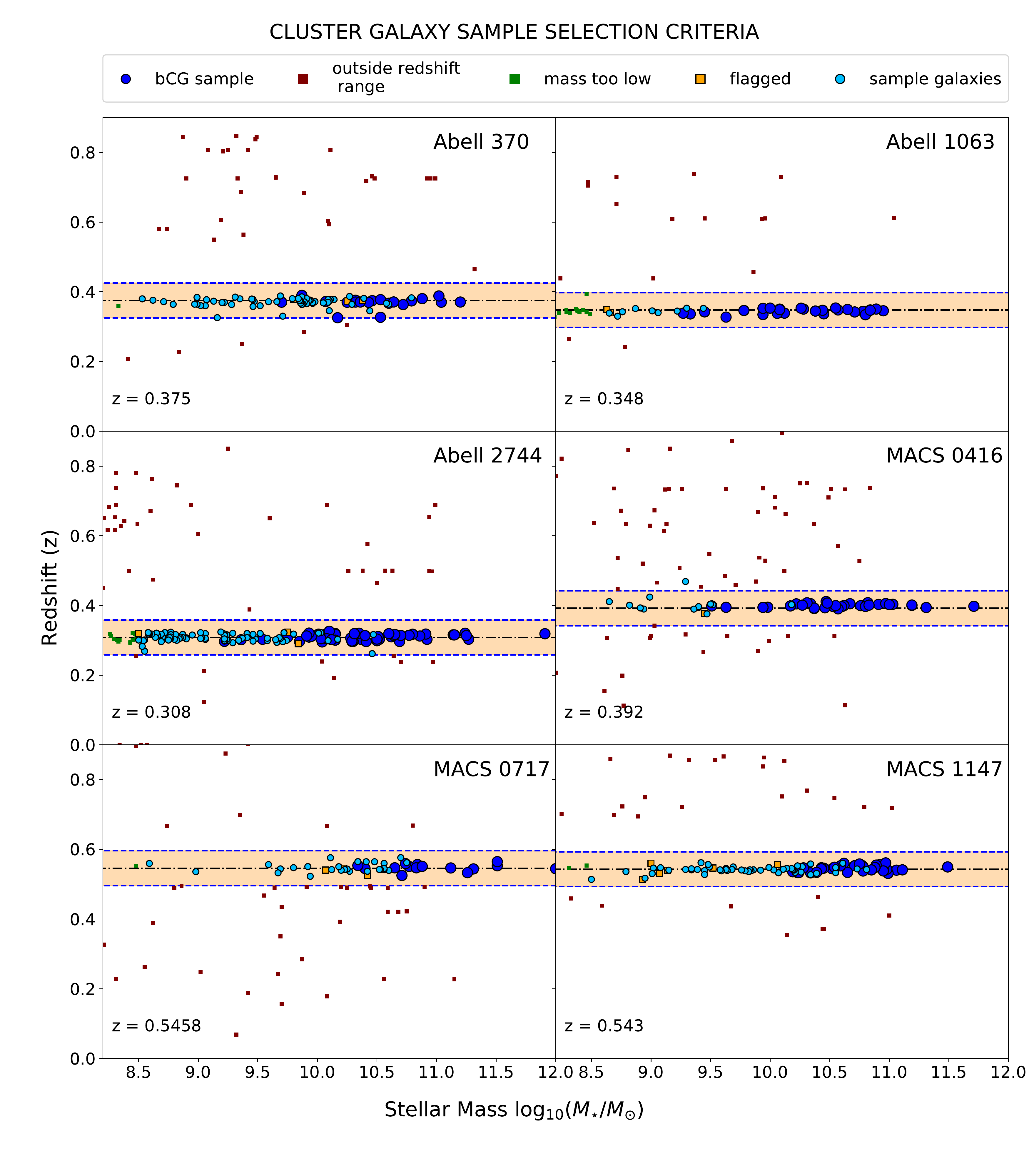}}
	\caption[Stellar mass and redshift distribution of the cluster sample.]{Plots showing redshift vs. stellar mass, demonstrating how the cluster sample of galaxies was selected. Black dot-dashed line shows where cluster redshift is, and blue dashed lines show $z_{cluster} \pm0.05$ redshift range where the cluster sample galaxies will be selected from. Only objects with spectroscopic redshifts are plotted, which is 3-5\% of the entire catalog. ``Flagged" objects are ones where their aperture is overlapping a masked region, or they have any bad flux/error/weight value (i.e. negative, NaN/Inf) for pixels associated with a source in the segmentation map.  }\label{fig:clu-sample}
\end{figure*}
\begin{figure*}
	\centerline{\includegraphics[width=0.8\textwidth]{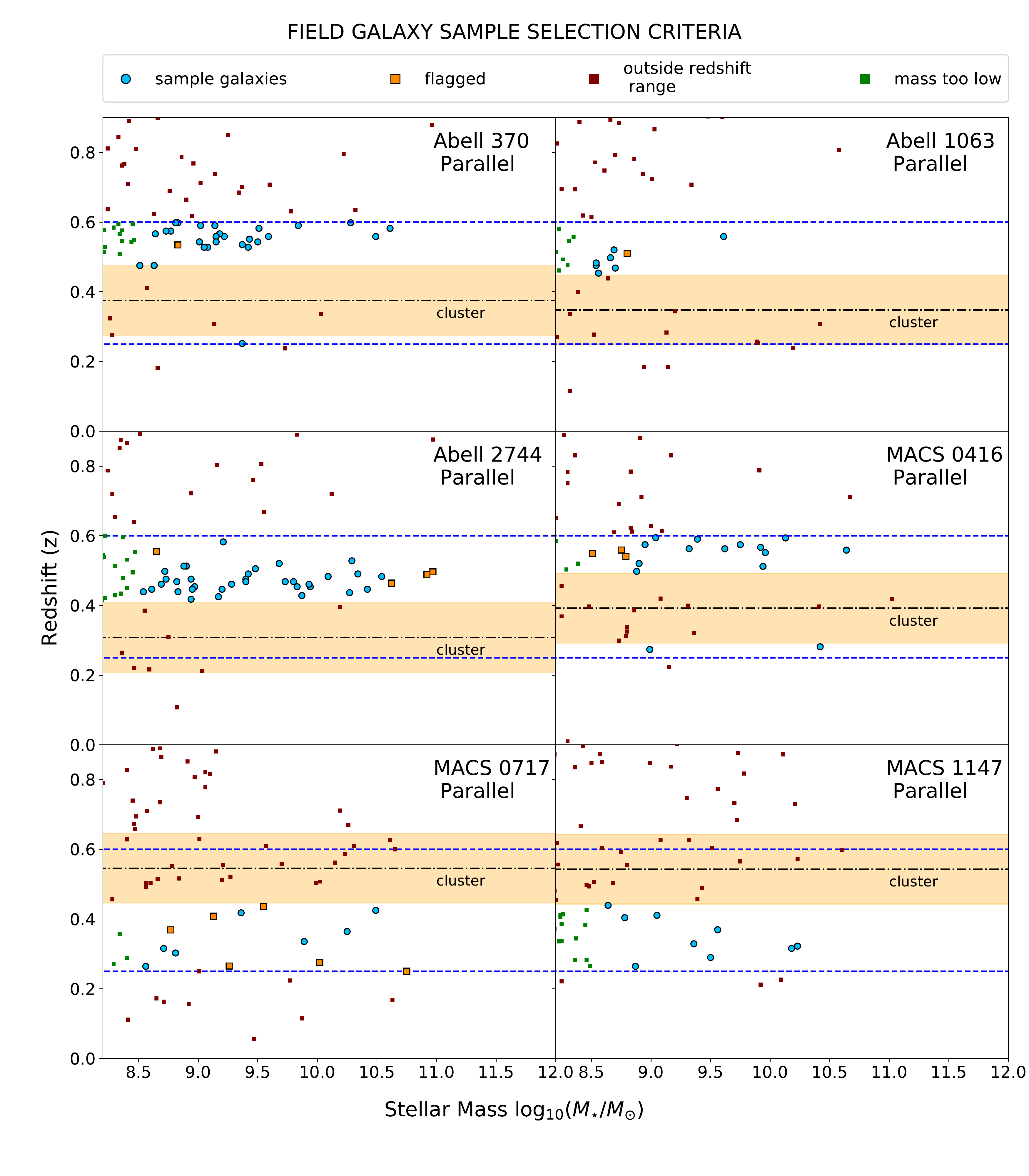}}
	\caption[Stellar mass and redshift distribution of the flanking fields sample.]{Plots showing redshift vs. stellar mass demonstrating how the field sample of galaxies was selected. The vast majority of the objects in the parallel fields do not have spectroscopically confirmed redshift so the exclusion range around the cluster is $z_{cluster} \pm0.1$ instead.  $z_{phot}$ values are plotted when the object has no $z_{spec}$ value. Selected galaxies are indicated by the light blue circles. Flagged objects follow the same rules as applied to the cluster sample. Blue dashed lines indicate the redshift boundaries for the field sample ($0.25 < z < 0.6$). }\label{fig:par-sample}
\end{figure*}


\subsubsection{Cluster sample}\label{sec:data-clu}
It is important to have a pure cluster sample for determining the environmental effects of quenching. For the cluster sample, only objects with spectroscopically-confirmed redshifts are included. Spectroscopic selection is essential because for objects that have both a $z_{phot}$ and a $z_{spec}$, 17\% of them have a difference of $>20\%$ between the two redshifts. In addition, 44\% of these objects have $z_{phot}$ measurements that are more than 0.05 off from their $z_{spec}$ measurements. Given that the typical velocity dispersion of the HFF clusters implies a redshift range of $\pm$ 0.05 - 0.08 for members, this means photometric redshifts are not precise enough to definitively determine which objects are cluster members. In order to minimize the interloper fraction we make a redshift cut for each cluster to only include galaxies that are within $ \pm 0.05$ of the cluster's redshift to make sure that only objects within the cluster's virial radius are selected. The six clusters each have a virial mass from $\sim 1 \times 10^{15} \msun$ to $\sim 3\times10^{15}\msun$. The $R_{200}$ of the clusters are between 1.89 -- 2.57 Mpc (See Table \ref{table:R200} and \citealt{L2017} for more details.)

For the clusters, we also include the bCGs in our sample to add more objects with higher stellar mass, in the range of $10^{10} - 10^{11.5} \msun$. Since these galaxies were subtracted from the original images before the detection image was created, their total flux was not determined using the segmentation map created by \textsc{SExtractor}, but through the process of modelling with isophotes as described in \S \ref{sec:data1}. Their total integrated flux is listed in the HFF Deep Space catalogs along with the non-bCG galaxies' fluxes.

Figure \ref{fig:clu-sample} displays all of the galaxies that were selected to be in the cluster sample as light blue circles. Any objects flagged according to the catalog are removed, and displayed in Figure \ref{fig:clu-sample} as other symbols. Redshift limits are marked with dashed blue lines that represent $z_{cluster}\pm 0.05$. After applying our selection criteria, the number of cluster galaxies in our sample is 400.


\subsubsection{Field sample}\label{sec:data-par}
For the field sample, precise redshifts are not a necessary requirement for selection of galaxies like in the cluster sample. Therefore, selecting a sample based on photometric redshifts is sufficient. The flanking/parallel fields also do not have many objects with spectroscopically confirmed redshifts. If we excluded all objects without a confirmed $z_{spec}$, the sample size would be too small for proper statistical analysis (our parallel field sample would comprise of only around 40 galaxies in total). Due to the small number of spectroscopically confirmed redshifts, we substituted the photometric redshift ($z_{phot}$) for any objects that do not have a $z_{spec}$ measurement. The photometric redshifts for the Frontier Fields objects were obtained via the code \textsc{EAzY} \citep{B2008}. The redshift range chosen for selecting the field sample is $0.25 \leq z \leq 0.6$. However, the galaxies with redshifts within $\pm$ 0.10 of the cluster's redshift are excluded. We omit a wider redshift range around the clusters in the parallel fields to be conservative because $z_{phot}$ measurements are less precise. Nevertheless, the parallel fields will likely still contain some cluster members if we include galaxies with $z_{phot}$ measurements close to the clusters' redshifts, because the distance from the cluster center to the center of their respective parallel fields is approximately the $R_{200}$ of the clusters. Using a wider redshift cut around the clusters ensures the low-density field sample does not have any contamination from cluster members, giving a ``true" field sample. 

Figure \ref{fig:par-sample} is a redshift vs. stellar mass plot similar to Figure \ref{fig:clu-sample}, but it does not show all of the unselected objects due to size of the catalog. 10\% of the full catalog's unselected objects are plotted in the Figure. The selected objects are however, displayed in their entirety. As mentioned in \ref{sec:data1},  certain galaxies in the parallel field images were also flagged as bCGs according to their sizes or because their light profiles are affected by a number of nearby galaxies. Our field sample contains 96 galaxies after applying all selection criteria.

\begin{figure*}
	\centerline{\includegraphics[width=0.7\textwidth]{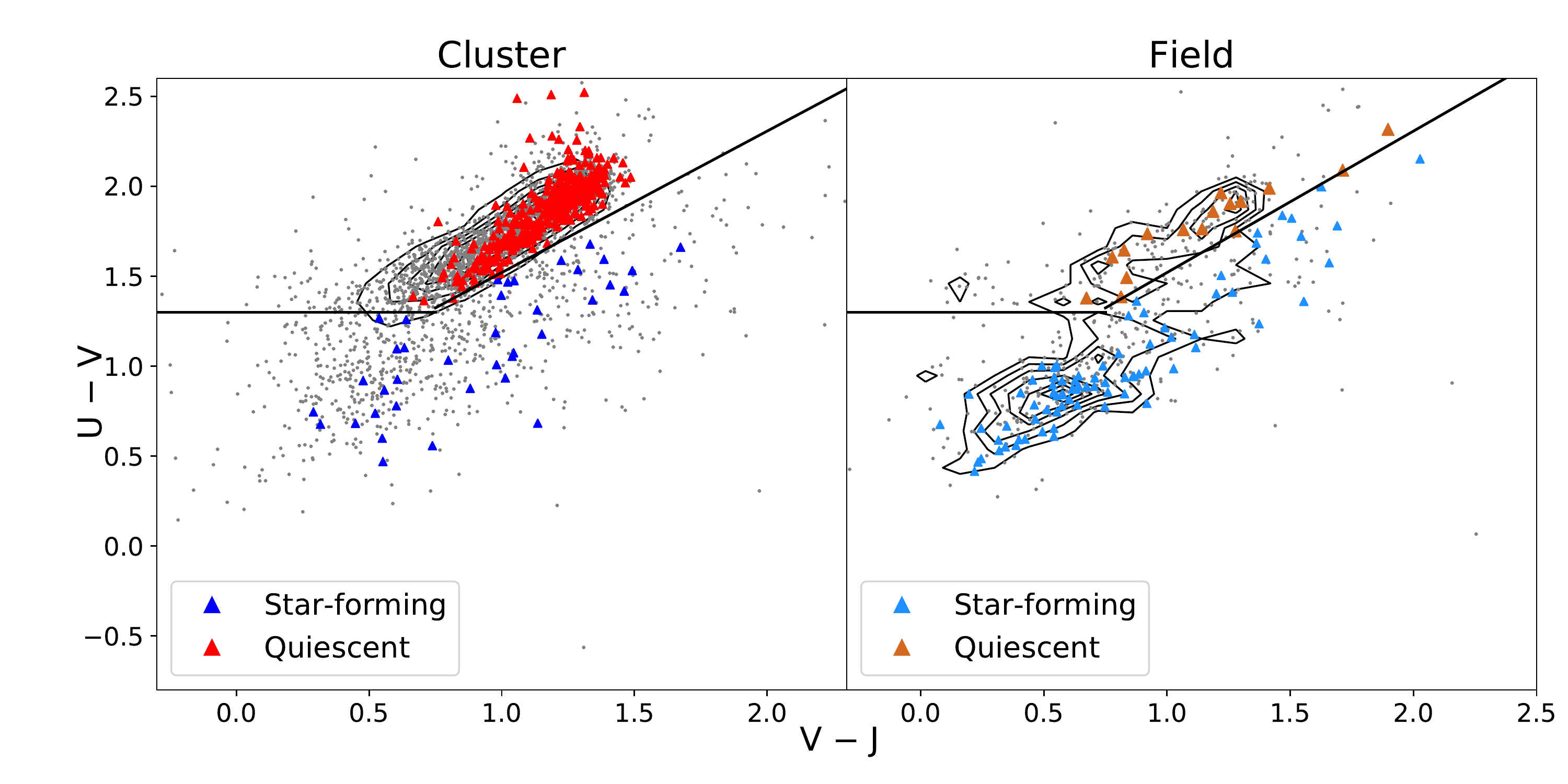}}
	\caption[UVJ diagram to separate star-forming and quiescent galaxies]{ The UVJ diagram for cluster galaxies (left) and field galaxies (right). The contours show U-V vs. V-J for the entire catalog, while the coloured points are the galaxies that made it into our final sample. Red and orange points are the quiescent galaxies, and the blue and cyan points are the star-forming galaxies for cluster and field, respectively. The boundaries for the UVJ diagram are from \cite{S2018}. \label{fig:uvj1}}
\end{figure*}

\fontsize{9}{9}\selectfont
\begin{threeparttable}
\begin{tabular}{@{}llll@{}}
\toprule
Cluster    & $R_{200}$ & angular sep. & separation \\
           & (Mpc)     & (arcsec)   & (Mpc)      \\ 
\midrule
Abell 370  & 2.57\tnote{a}& 354.35"    & 1.83       \\
Abell 1063 & 2.55\tnote{b}& 360.31"    & 1.77       \\
Abell 2744 & 2.40\tnote{c}& 361.44"    & 1.64       \\
MACS 0416  & 1.89\tnote{d}& 358.20"    & 1.91       \\
MACS 0717  & 2.36\tnote{d}& 359.99"    & 2.30       \\
MACS 1149  & 2.35\tnote{d}& 360.54"    & 2.30       \\ 
\bottomrule
\end{tabular}
\begin{tablenotes}
\item[a] \cite{Lah:2009}
\item[b] \cite{Zenteno:2016}
\item[c] \cite{Boschin:2006}
\item{d} \cite{Umetsu:2016}
\end{tablenotes}
\caption{Comparison of virial radius $R_{200}$ of each cluster with the separation of the cluster center from the parallel field center. The angular separation is obtained from the J2000 coordinates of cluster center and parallel field center given in \cite{L2017}.\label{table:R200}}
\end{threeparttable}

\normalsize 

\subsubsection{Dividing the Cluster and Field samples by Star-formation activity}\label{sec:data-uvj}

The cluster and field samples are further divided into quiescent and star-forming based on their position on the UVJ diagram \citep{Wi2009} in Figure \ref{fig:uvj1}. The boundaries between star-forming and quiescent for the UVJ diagram are the same as from S18, which are defined by the equations
\begin{linenomath*}
\begin{align}
    U-V< 1.3 \text{ for } V-J< 0.75 \label{eqn:uvjbound1}  \;\;,\\
    U-V< 0.8(V - J) + 0.7 \text{ for }(V-J) \geq 0.75 \label{eqn:uvjbound2} \;\;.
\end{align}
\end{linenomath*}
The reason for dividing the sample into different types defined by both their environment and their star-formation rate is for two purposes. Firstly, it shows how environment on its own affects the percentage of galaxies that are quiescent. Secondly, we quantify similarity in morphology through similarity in 2D profile fitting parameters. This is to link progenitors in the field with descendants in the cluster in a statistical way, which is based on the assumption that the distribution of stellar mass is conserved as galaxies undergo environmental quenching. 

\begin{figure*}
	\centerline{\includegraphics[width=0.8\textwidth]{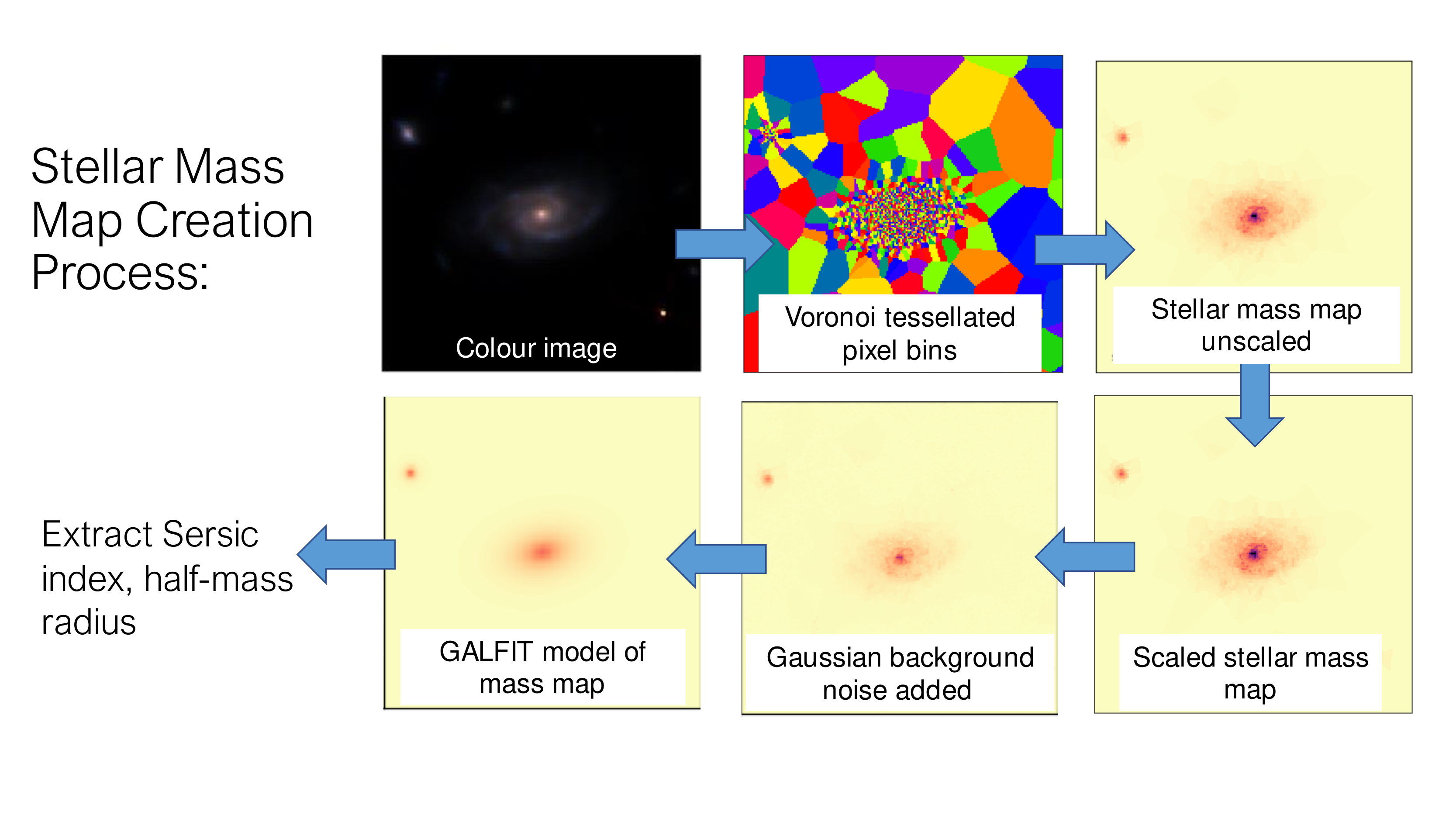}}
	\caption{The summary of the steps for constructing the resolved stellar mass maps for one cluster galaxy.}\label{fig:process}
\end{figure*}

Panels in Figure \ref{fig:uvj1} show the UVJ colour diagram where the star-forming and quiescent galaxy samples are defined for both clusters (left panel) and field (right panel). The coloured points correspond to the colour distributions of the galaxies in our selected cluster member and field samples. There are significantly more quiescent galaxies in the clusters (red points) than in the field (orange points), and likewise, more star-forming galaxies in the field (cyan points) than in the cluster (blue points). We will be focusing on how the star-formation activity and the environment both affect the morphology of our galaxy sample in the analysis in \S\ref{chap:results}.

\section{Resolved stellar mass maps}\label{chap:method}

  Constructing a stellar mass map is the most direct way to probe how stars are distributed within a galaxy. This makes them useful for comparing star-forming to quiescent galaxies of similar stellar mass. Figure \ref{fig:process} shows the steps taken to construct a resolved stellar mass map in order to quantify the stellar mass morphology of our galaxy sample. When galaxies evolve off the star-forming main sequence and become quiescent, stellar mass is likely to be conserved, in the assumption that quenching is not due to a merger. This is because quenching can potentially mean no more gas is available to be turned into stars, so the mass of stars is effectively  ``frozen". Morphology for galaxies can be arbitrary and subjective, but for this study, since stellar mass distributions can also be modelled by S\'ersic profiles \citep{S1963}, the parameters of the S\'ersic models of stellar mass maps can serve as a way of quantifying morphology.
 
 We present our methods for producing resolved stellar mass maps of the selected galaxy sample, and the single-component 2D S\'ersic profiles derived from the stellar mass. Cutout images of each galaxy were taken in each available band for the cluster with the dimensions of the postage stamp being around 15 to 20 times the half-light radius listed in the HFF Deepspace catalog (which was obtained via SExtractor). Similar sized cutouts of the segmentation map were made as well for the purposes of distinguishing object and background pixels. 


\subsection{Resolved SED-fitting}\label{sec:method-SEDfit}
The spectral energy distribution (SED) of a galaxy informs its mass-to-light ratio ($M/L$), which changes as the galaxy evolves, since older stellar populations are less luminous and output more in the IR than in the UV. This means the total stellar mass of a galaxy can be determined from its SED. High-quality spectra are not always available, especially for higher redshifts as is the case for our sample of galaxies, so stellar mass can also be obtained by fitting SEDs of various stellar populations to photometry in the absence of spectra. 

Most of the work on SED-fitting of distant galaxies is from ground-based observations, which means the galaxies do not have enough resolution for detailed spatial analysis of the stellar mass, and only the total unresolved stellar mass of the entire galaxy can be inferred. With the HFF HST data, we have enough angular resolution to map the stellar mass distribution in two spatial dimensions with multiband data. The fit becomes more accurate with more photometric bands, and as stated in \S \ref{sec:data1}, there are 7-14 HST bands utilized for photometry.

The brighter parts of the galaxy have high signal to noise (S/N). However, the outer regions begin to approach the sky noise, and hence our modeling of the galaxy needs to account for the large range in S/N for pixels in the galaxy. If SED-fitting was performed pixel by pixel, the SED and measured stellar mass would be highly uncertain, especially for the outskirts of galaxies, or galaxies that have lower surface brightness. 

Pixel binning can fix this problem by creating bins with similar signal-to-noise ratios. The papers by \cite{W2012},  \cite{L2014}, and \cite{C2016} all describe various methods of stellar mass map construction by SED-fitting to photometry. Here we employ our own methodology similar to the ones described in the above works but with modifications specific to our dataset. 

\subsubsection{Pixel binning for maximizing signal-to-noise in galaxy subregions}\label{subsec:method-binning}
This work uses the Voronoi binning technique created by \cite{CC2003}, which was also utilized in \cite{W2012}. Voronoi tessellation is most effective when modelling galaxies that have a large range in S/N per pixel. This binning method creates spatial bins of equal S/N in order to best homogenize the wide range in S/N.  

\begin{figure}
	\centering
	\begin{tabular}{cc}
	    F160W band image \hspace{1.3cm} Voronoi tessellated bins\\
		\includegraphics[width=4cm]{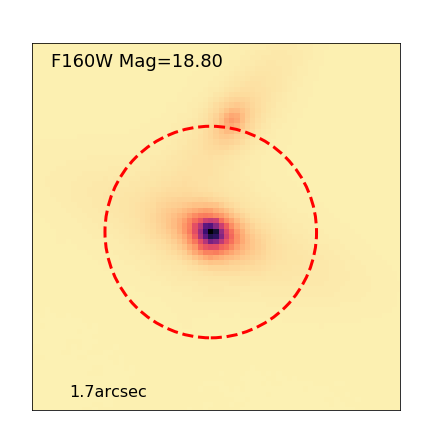}
		\includegraphics[width=4cm]{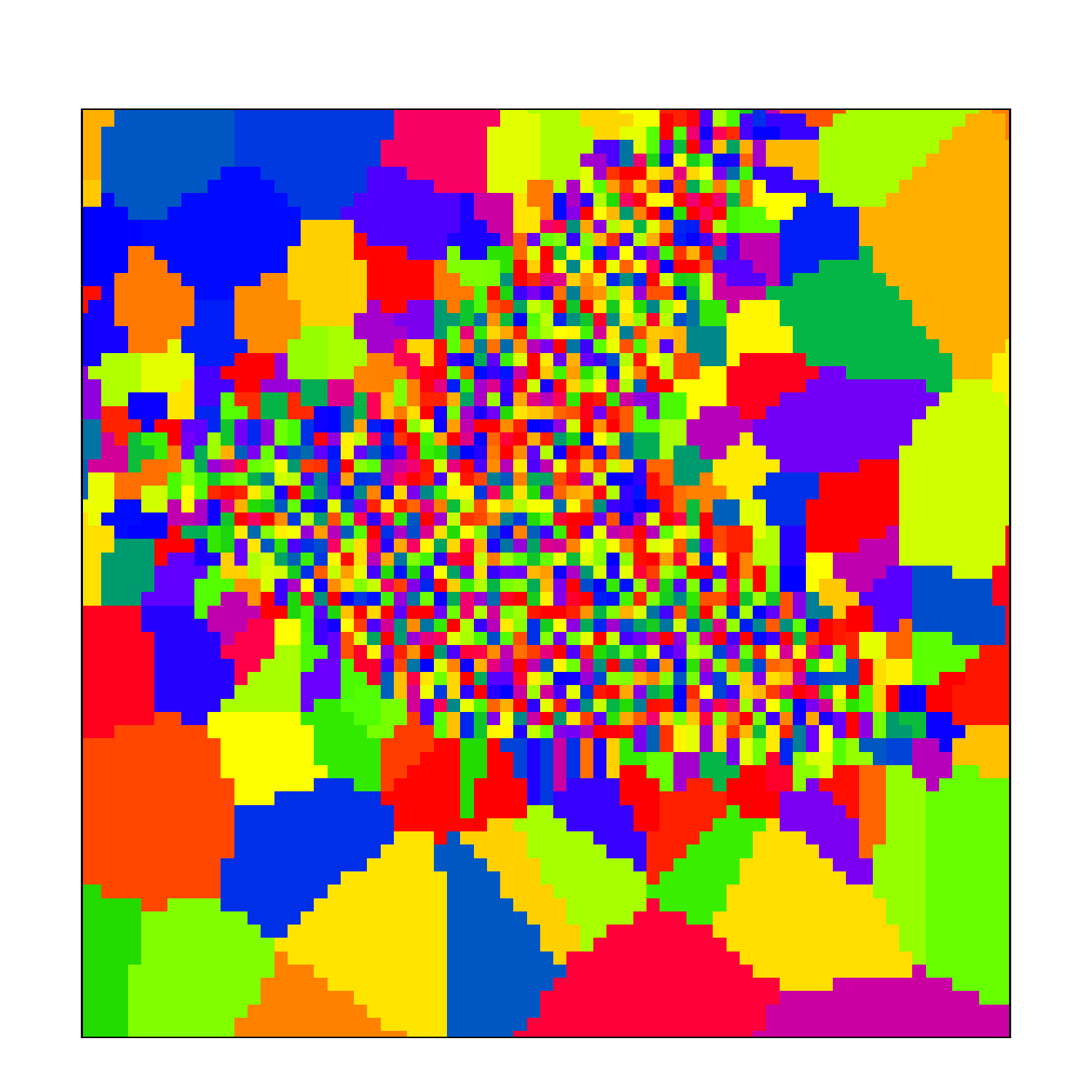}\\
		\includegraphics[width=4cm]{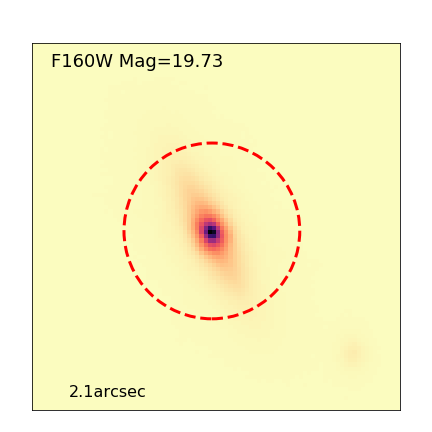}
    	\includegraphics[width=4cm]{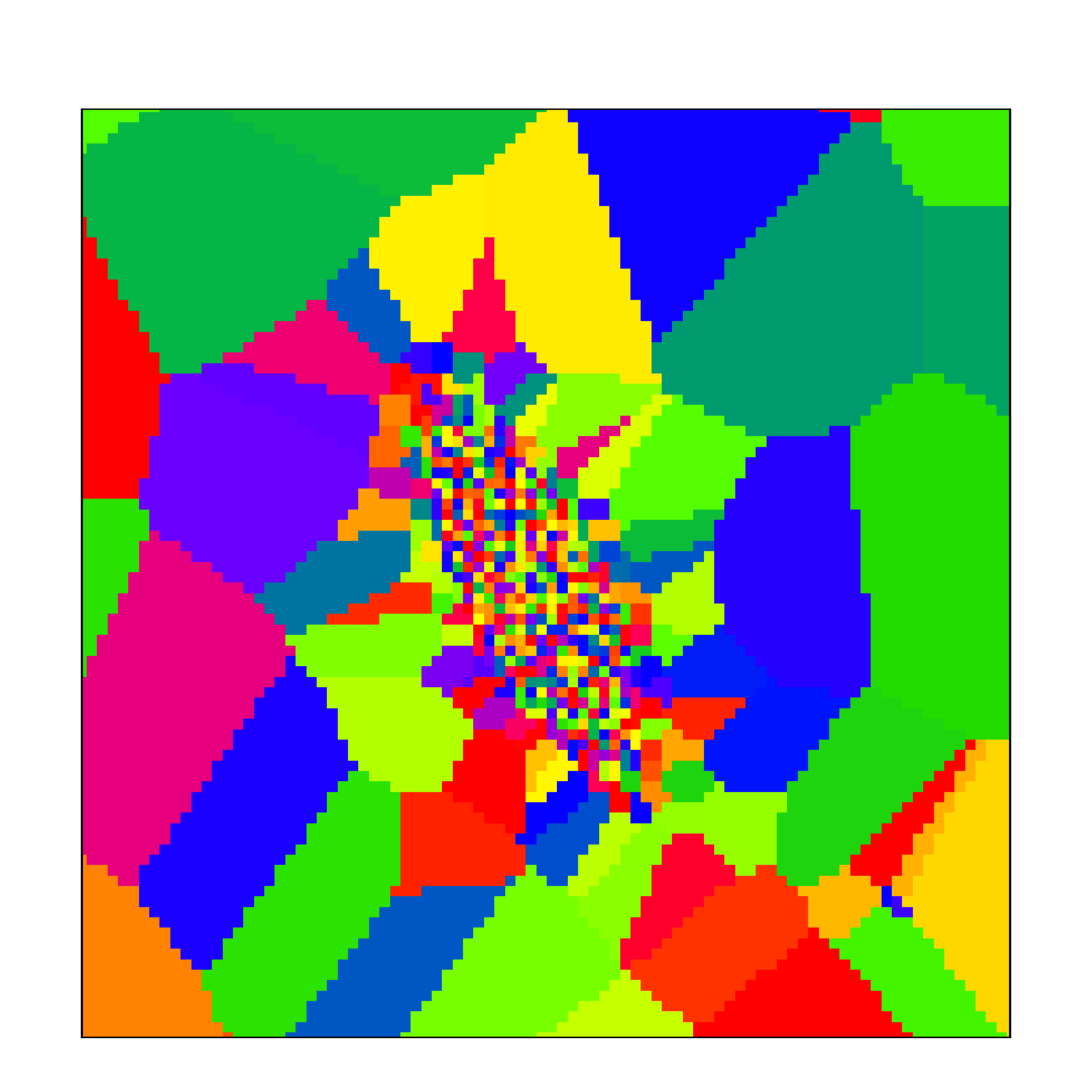}\\
		\includegraphics[width=4cm]{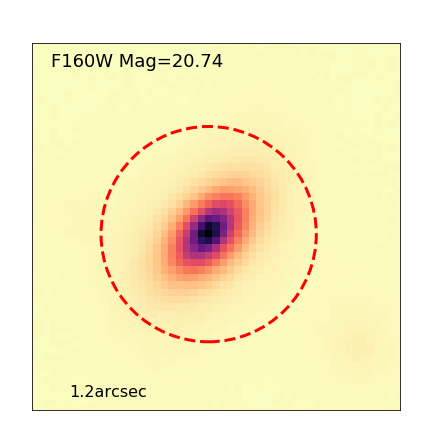}
		\includegraphics[width=4cm]{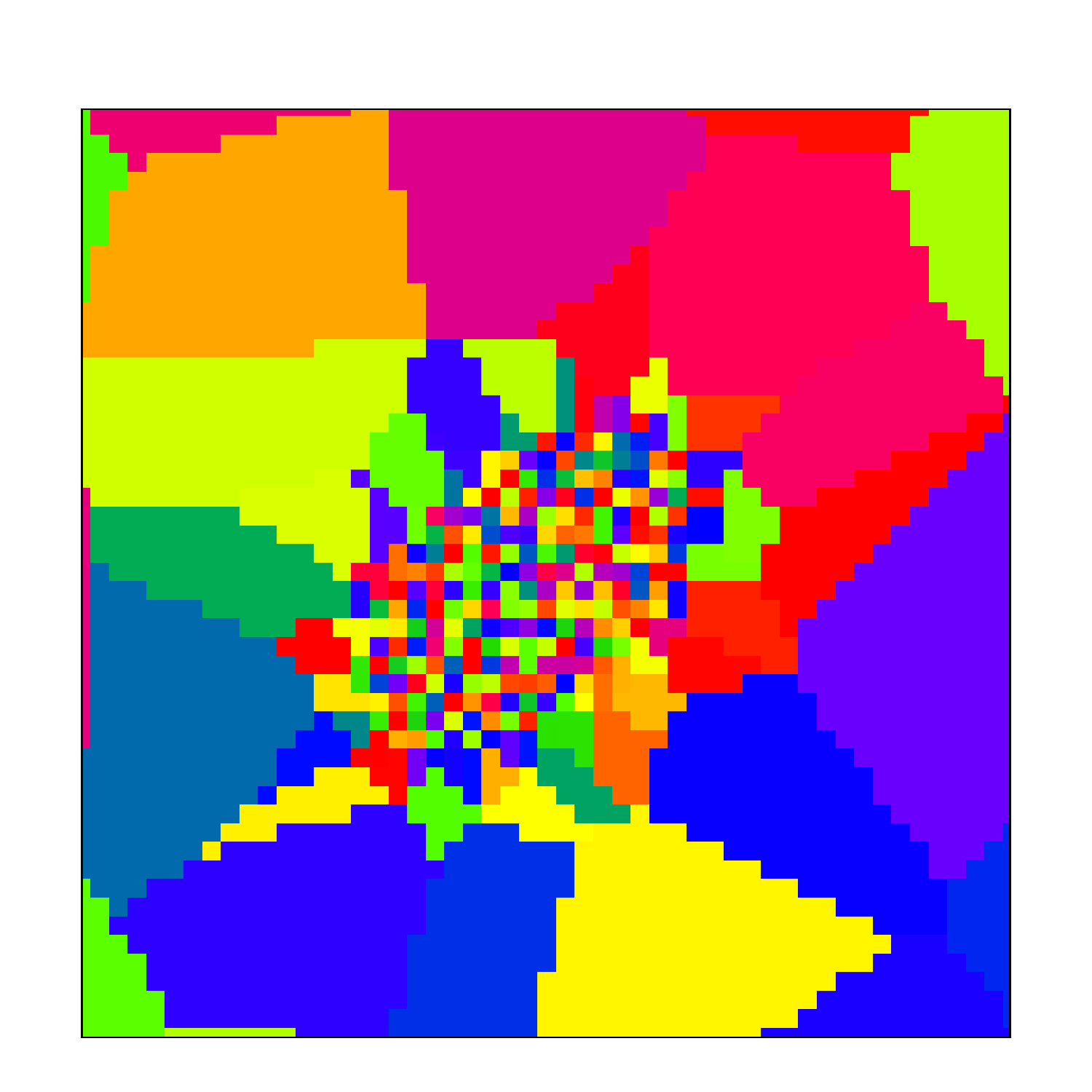} \\
	\end{tabular}
\caption{ Examples of pixel binning using the weighted Voronoi tessellation algorithm from \cite{CC2003}. F160W flux is the basis for the binning since IR flux correlates well to stellar mass distributions. The red circle denotes 3 times the F160W effective radius of each galaxy.\label{fig:binning-example}} 
\end{figure}

\cite{CC2003} designed two algorithms, weighted Voronoi tessellation (WVT) and centroid Voronoi tessellation (CVT), which work in tandem to create bins that are as circular as possible by minimizing the distances between each centroid. However, the WVT was not designed to work well with images that contain a large background with low S/N. It compensates for the low flux bins' lack of signal by adding unconnected high S/N pixels from inside the galaxy, which would result in rings  of pixels from inner regions of galaxies getting added onto background bins. Only using CVT alone fixes this issue, although this results in sharper bins. This change in bin shape does not result in a loss of detail as bins in regions with high S/N remain small (see Figure \ref{fig:binning-example}, right column). Since only one filter can be used to determine bin placement, we chose the F160W photometry as the basis for the algorithm to construct the spatial bins, since IR flux correlates well to stellar mass distributions.

For most galaxies, the minimum S/N for a single bin depends on the flux of the background pixels in relation to the target object's pixels. Galaxies with less flux also tend to have a smaller angular size. The number of bins are checked against the size of the image to ensure that they have a minimum of 100 spatial bins for adequate spatial resolution before SED-fitting. If the minimum is not satisfied, the target S/N for each spatial bin is decreased. If the flux in the target galaxy's pixels is high, the S/N for each bin will increase accordingly, but we limit the S/N for each bin to not exceed a value of 30. The lower limit for most galaxies with low flux in the F160W filter is a S/N of 10, although for certain low surface brightness galaxies, the binning would fail unless the S/N for each bin was set to 5. What particular S/N is selected for a certain image also depends on the size of the image in relation to the number of bins. 
In Figure \ref{fig:binning-example} we show three examples of how the pixel binning algorithm generates a bin map. Pixel binning will result in edges of bins leaving a residual on the resolution, and the effects of those edges means there must be a lower limit for the number of bins in an image. Once pixel binning has finished and every galaxy has the total flux of each of its bins in every band catalogued in a data table, this catalog becomes the input for the resolved. SED-fitting.

\subsubsection{Constructing the stellar mass map}\label{subsec:method-FAST2}

To create stellar mass maps we fit the 7-14 band photometry in each Voronoi bin to stellar population synthesis models. To do this we utilized \textsc{FAST} (Fitting and Assessment of Synthetic Templates, \citealt{K2009}). For SED-fitting, we use the stellar population synthesis (SPS) code from \cite{BC2003} (hereafter BC03), a Chabrier (\citeyear{C2003}) IMF, and a Calzetti dust law \citep{C2000}. Every bin in a galaxy's Voronoi spatial bin will use the same spectroscopic or photometric redshift as the galaxy itself, depending on whether $z_{spec}$ is available for that particular galaxy. The error in the goodness of fit of the SEDs are negligible between using $z_{spec}$ or $z_{phot}$ with FAST at this redshift range \citep{M2009}. A delayed tau model for the star formation history (SFH) is adopted, which is a parameterization where star formation rate is a function of time given by the equation
\begin{linenomath*}
 \begin{equation}
     \text{SFR}(t) \propto t e^{-t/\tau} \;\;,
 \end{equation}
 \end{linenomath*}
 where $\tau$ is a free parameter. This results in $\log(\text{SFR})$ initially increasing but then exponentially declining over time, with $\tau$ describing how fast it declines, and when the peak of the SFH is (the given range of possible values for $\tau$ is $6.5 \log(\text{yr}) < \log(\tau) < 11\log(\text{yr})$). The limits of integration for $t$ (age), and any other parameters were all set to their maximum limits (which is from $log(t) = 6\log(\text{yr})$ to $log(t) = 11\log(\text{yr})$) so that hitting the bounds of integration does not affect the computation of the stellar mass in the final output. FAST creates an output file that contains the stellar mass of each entry in units of $\log(\msun)$. 

\begin{figure*} 
	\centering
	\begin{tabular}{ccc}
		F160W flux & Stellar Mass Distribution  &Stellar Mass weighted \\
		& &by  F160W flux\\
		\includegraphics[width=5cm]{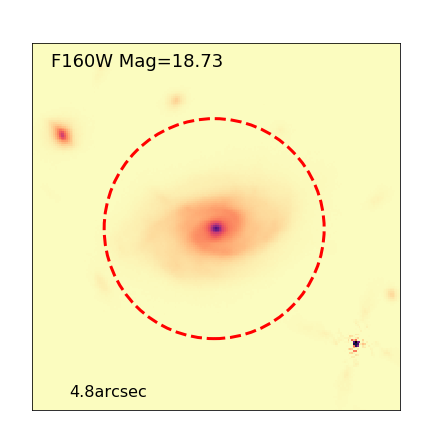}&
		\includegraphics[width=5cm]{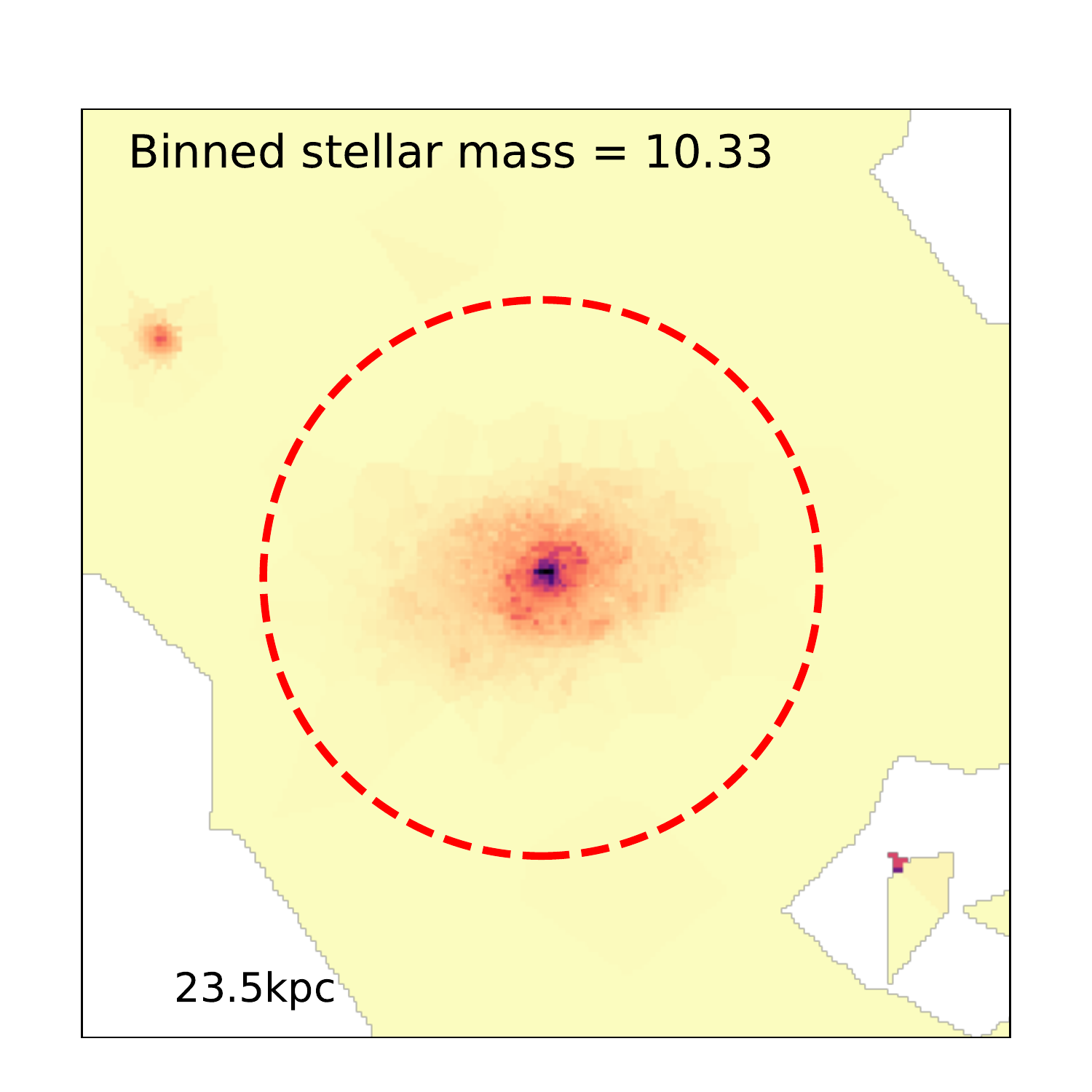}&
		\includegraphics[width=5cm]{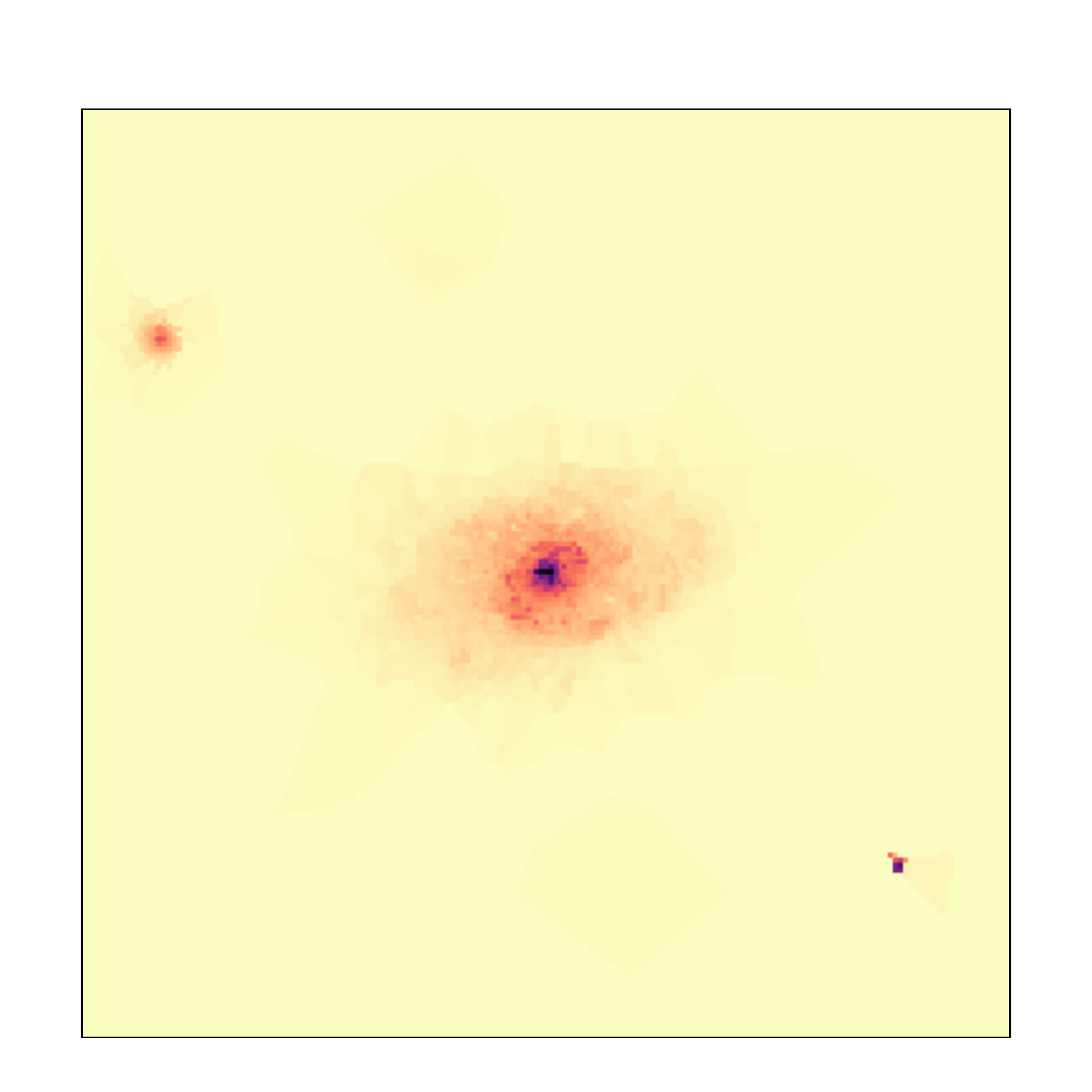} \\
		
		\includegraphics[width=5cm]{f160w_example40.png}&
		\includegraphics[width=5cm]{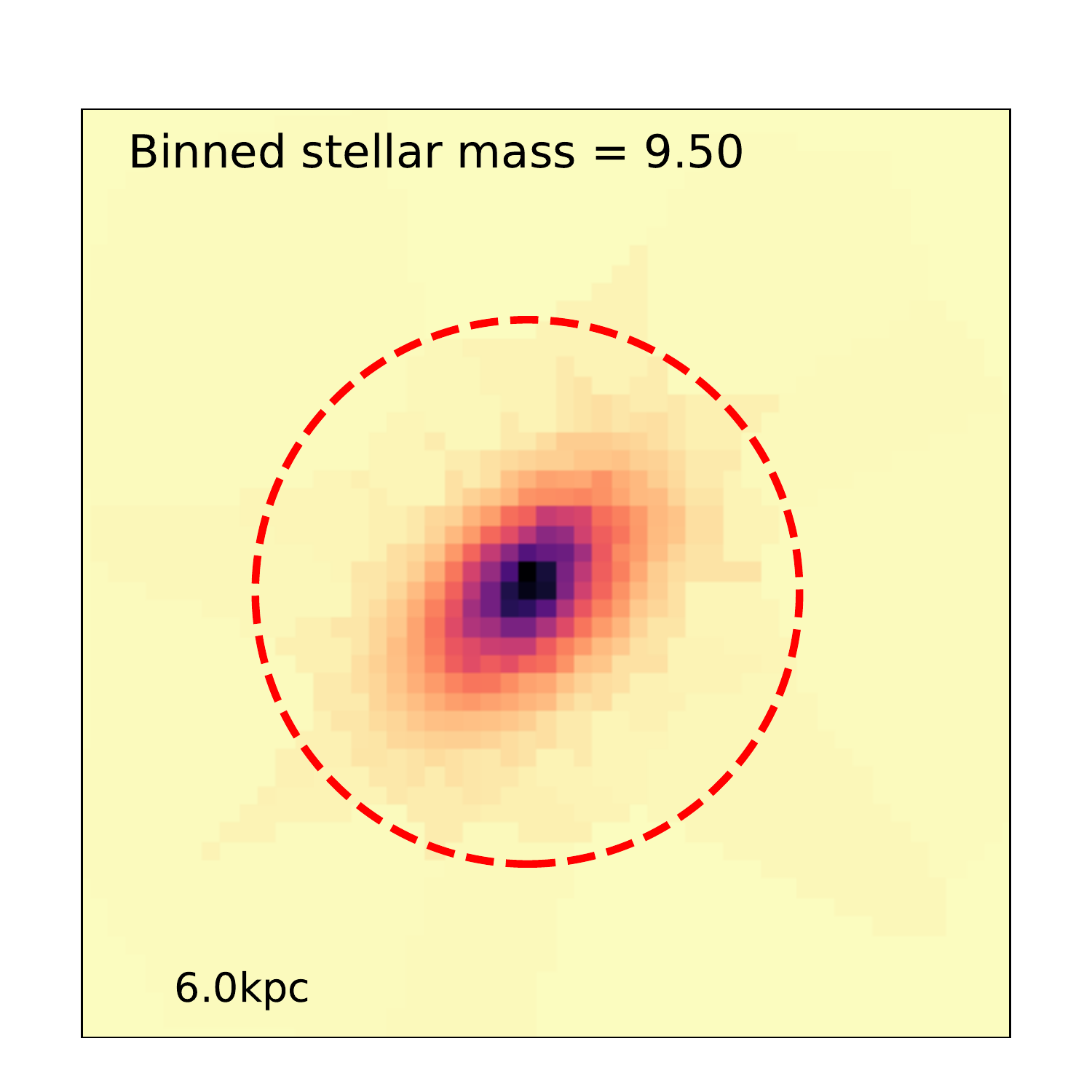}&
		\includegraphics[width=5cm]{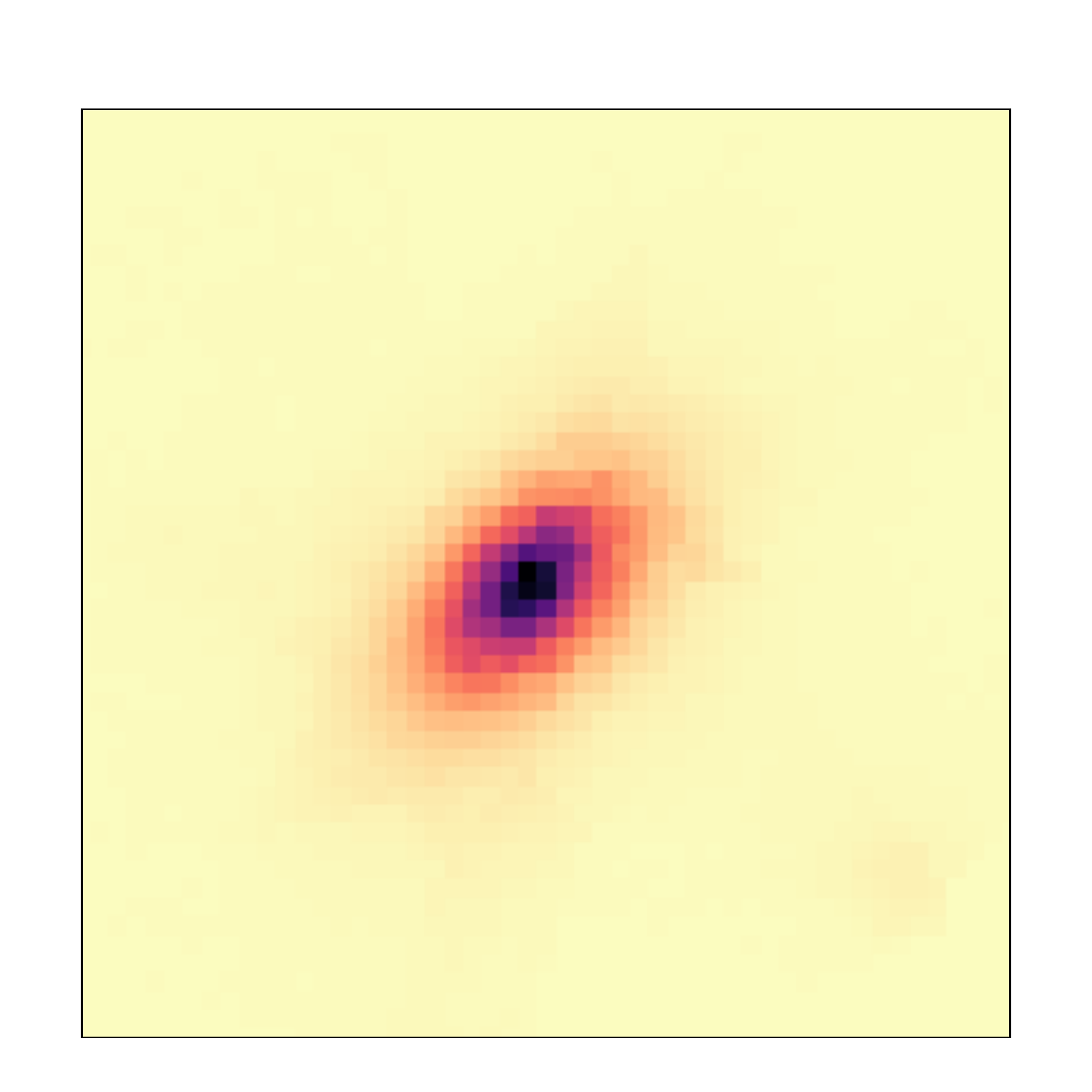} \\
		
		\includegraphics[width=5cm]{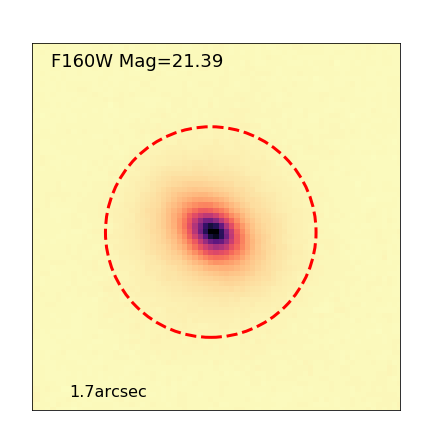}&
		\includegraphics[width=5cm]{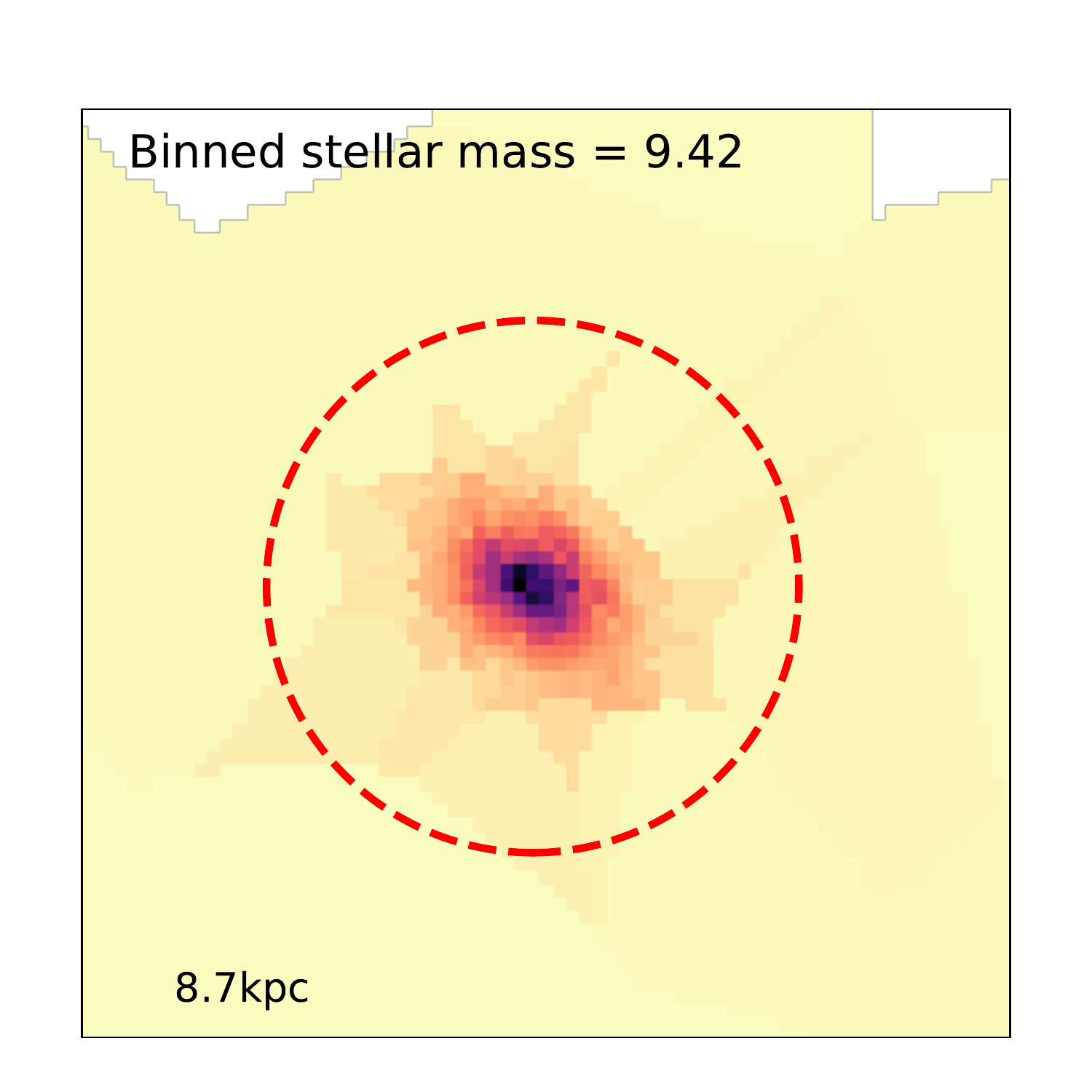}&
		\includegraphics[width=5cm]{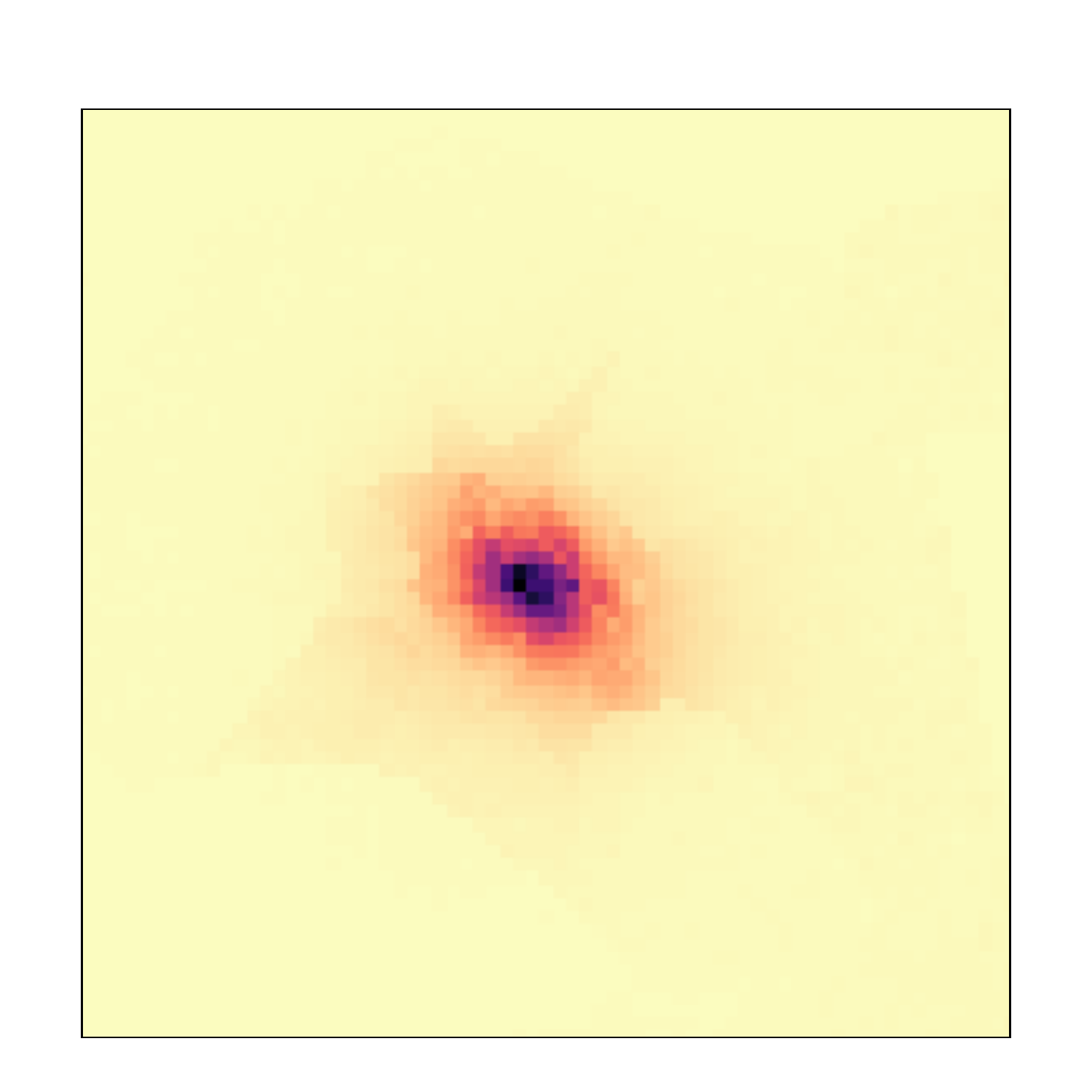}
	\end{tabular}
	\caption{Scaling the pixels in each bin helps recover some resolution. The red circle is 3$\times$ the effective radius of each galaxy as defined by SExtractor. Note the differences between the raw stellar mass outputs from FAST in the middle column and the pixel flux-weighted stellar mass maps in the right column are more pronounced for galaxies with less surface brightness (such as the galaxies in the middle and bottom row, where the flux-weighted stellar mass maps show a decrease in pixel-binning artefacts). The weighting is described in equation \ref{eqn:flux-weight}. The quoted values on the bottom left of panels reflect the effective radius for each object.}
	\label{fig:massmap-comp} 
\end{figure*}
 
A stellar mass map at the resolution of the spatial bins can be generated by assigning the bins in the image the stellar mass calculated by FAST divided by the number of pixels in the bin. However, each spatial bin is of a different size as they were generated by the Voronoi tessellation algorithm, such that high S/N areas have many bins of few or one pixel, while low S/N areas have few large spatial bins with many pixels. In order to recover some of the original resolution of the HST images, an extra step to this process is applied, where the percent flux contribution of each pixel to the overall flux of the bin is accounted for. This scaling can also be described in an equation as
\begin{linenomath*}
\begin{equation}
	M_{\text{pix,}i} =  \left( \frac{1}{ M_{bin} }\right) 
	\frac{F_{160}^{\text{pix},j}} {\sum_{\text{bin}}^{\text{pix},j} F_{160}^{\text{pix} } } \;\;, \label{eqn:flux-weight}
\end{equation}
\end{linenomath*}
where $M_{bin}$ is the mass FAST derived for the bin, $M_{\text{pix,}i} $ is the mass assigned to an individual pixel belonging to that bin, $F_{160}^{\text{pix},j}$ is the flux in the F160W band of that individual pixel, and the sum in the denominator is the total flux of the the entire bin. 

Figure \ref{fig:massmap-comp} demonstrates the difference that scaling by contributed flux makes to 
the bin map. The middle column shows the stellar mass distribution without scaling by percent flux contribution, and the right column is after the scaling. The stellar mass listed in the middle column refers to the total stellar mass of the galaxy summed from each pixel of the galaxy. There is significantly more information recovered for the latter maps that is lost when only using tessellated bins. The central regions of galaxies gain more angular resolution, and there is better distinction between which bins have more galaxy or background pixels, especially in more crowded images such as the second row of Figure \ref{fig:massmap-comp}. The F160W flux is closely correlated to stellar mass at these redshifts, so this percent scaling is reasonable, and the mass map becomes more representative of the actual distribution of stellar mass, with angular resolution consistent with the unbinned F160W HST images (Figure \ref{fig:massmap-comp}, left column). 

\begin{figure*} 
	\begin{subfigure}
		\centering
		\includegraphics[width=\columnwidth]{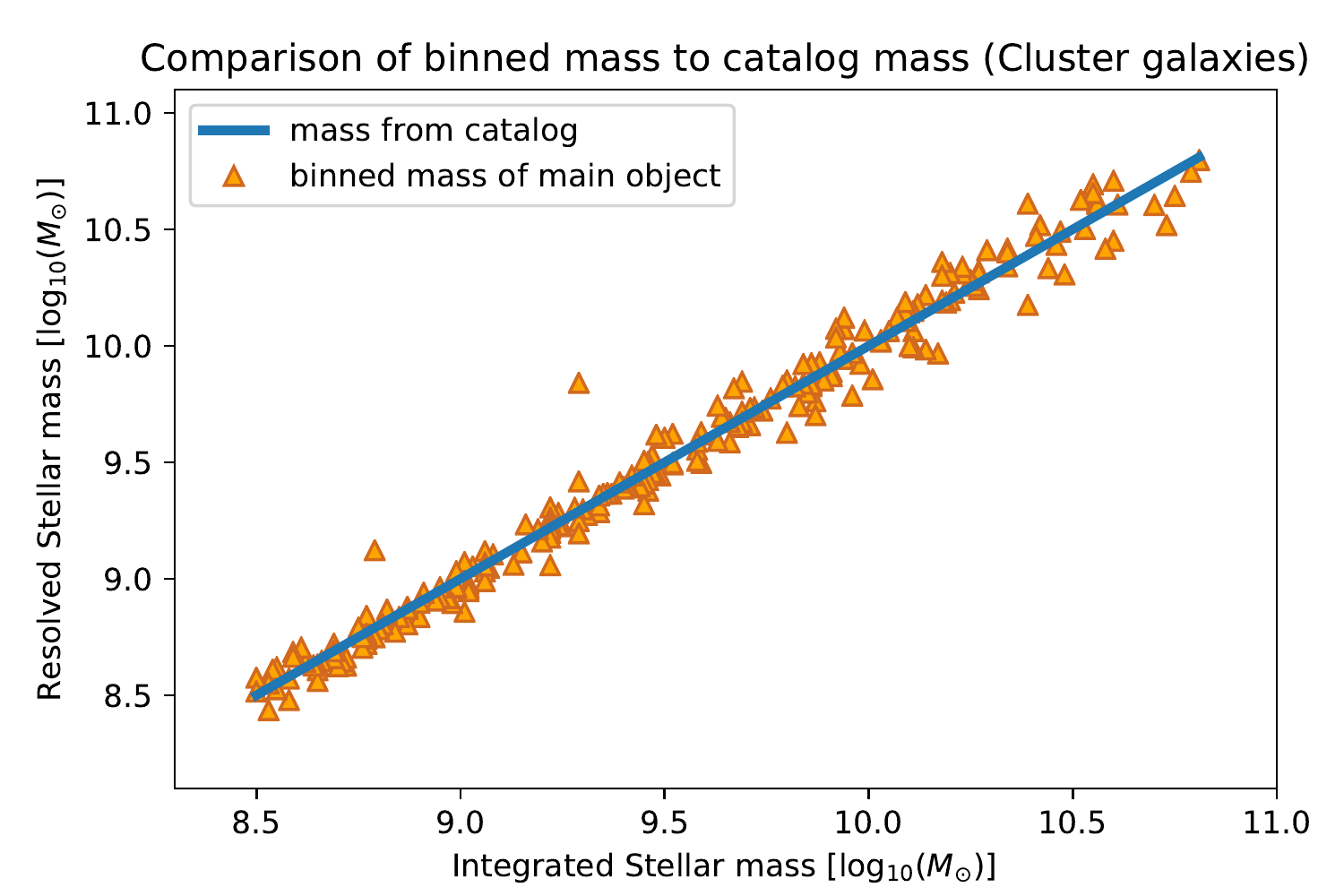}
	\end{subfigure}
	\begin{subfigure}
		\centering
		\includegraphics[width=\columnwidth]{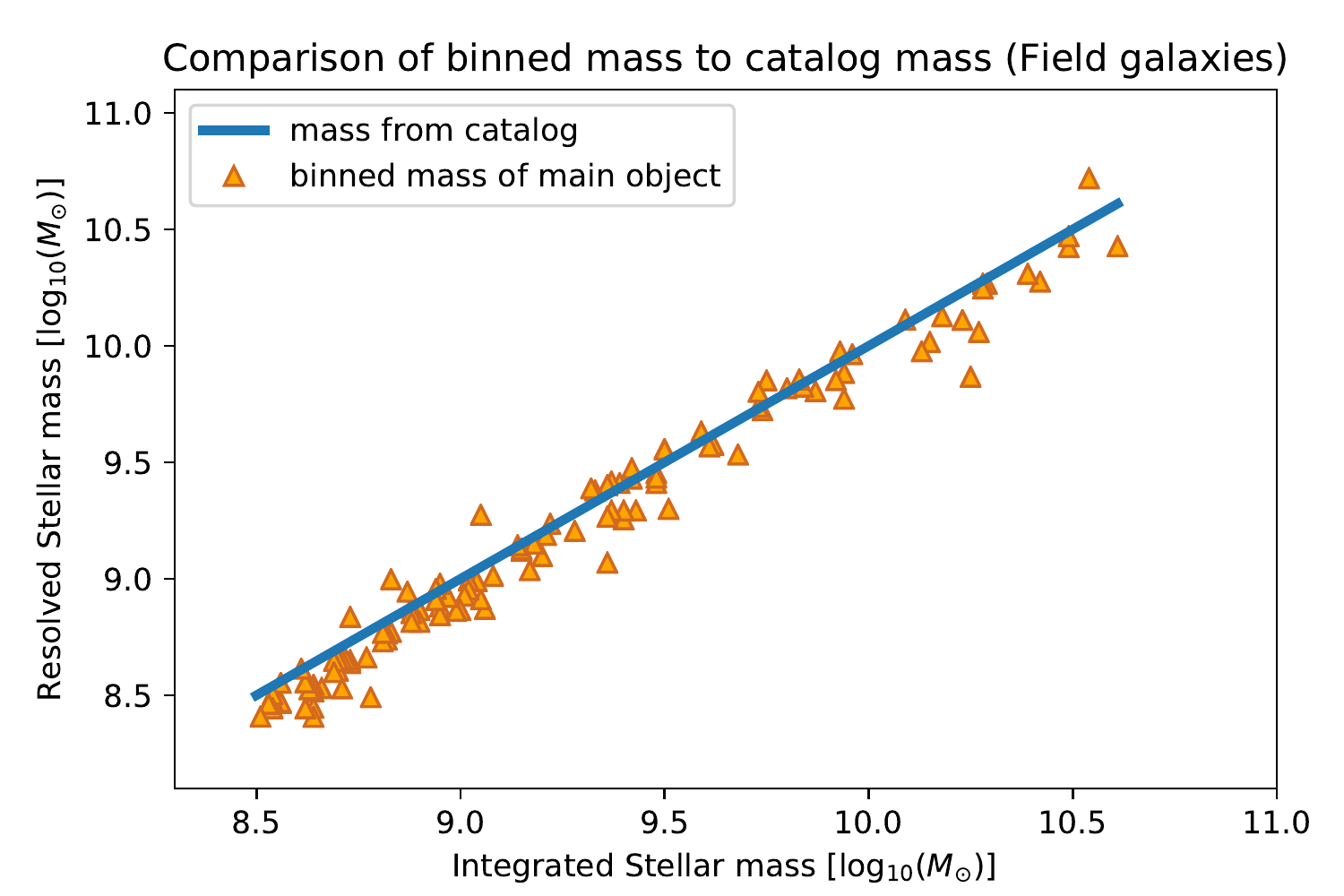}
	\end{subfigure}
	\caption{The plots compare the galaxies' total stellar mass calculated from resolved SED-fitting (orange triangles) against their integrated stellar mass from the HFF Deepspace catalogs (blue line). The x-axis is their integrated stellar mass from the catalog, and the y-axis is their resolved stellar mass. On the whole, the resolved stellar mass is lower by an average of 0.03 dex than the corresponding galaxy's stellar mass in the catalog.}	\label{fig:massplot}
\end{figure*}

An interesting byproduct of creating these resolved stellar mass maps is that in principle, the accuracy of the total stellar mass for each galaxy should be improved by resolved fitting (at the Voronoi bin level), as opposed to integrated fitting over the entire galaxy. In Figure \ref{fig:massplot}, the catalog's total stellar mass of each object is compared with the resolved total stellar mass obtained with our methods. The stellar mass from the HFF catalog was also derived with FAST, but over the entire flux of the galaxy instead of on individual pixel bins, hence it is the ``unresolved" mass. Note that although the two masses almost have a one-to-one relation, the resolved total mass is on average 0.03 dex lower, which is much less than the dispersion around the one-to-one in Figure \ref{fig:massplot}. \cite{SS2015} found similar results in their work on stellar mass from SED-fitting for the local universe with SDSS data.
\begin{figure*} 
\centering
	\begin{subfigure}
	\centering
		\includegraphics[width=0.75\textwidth]{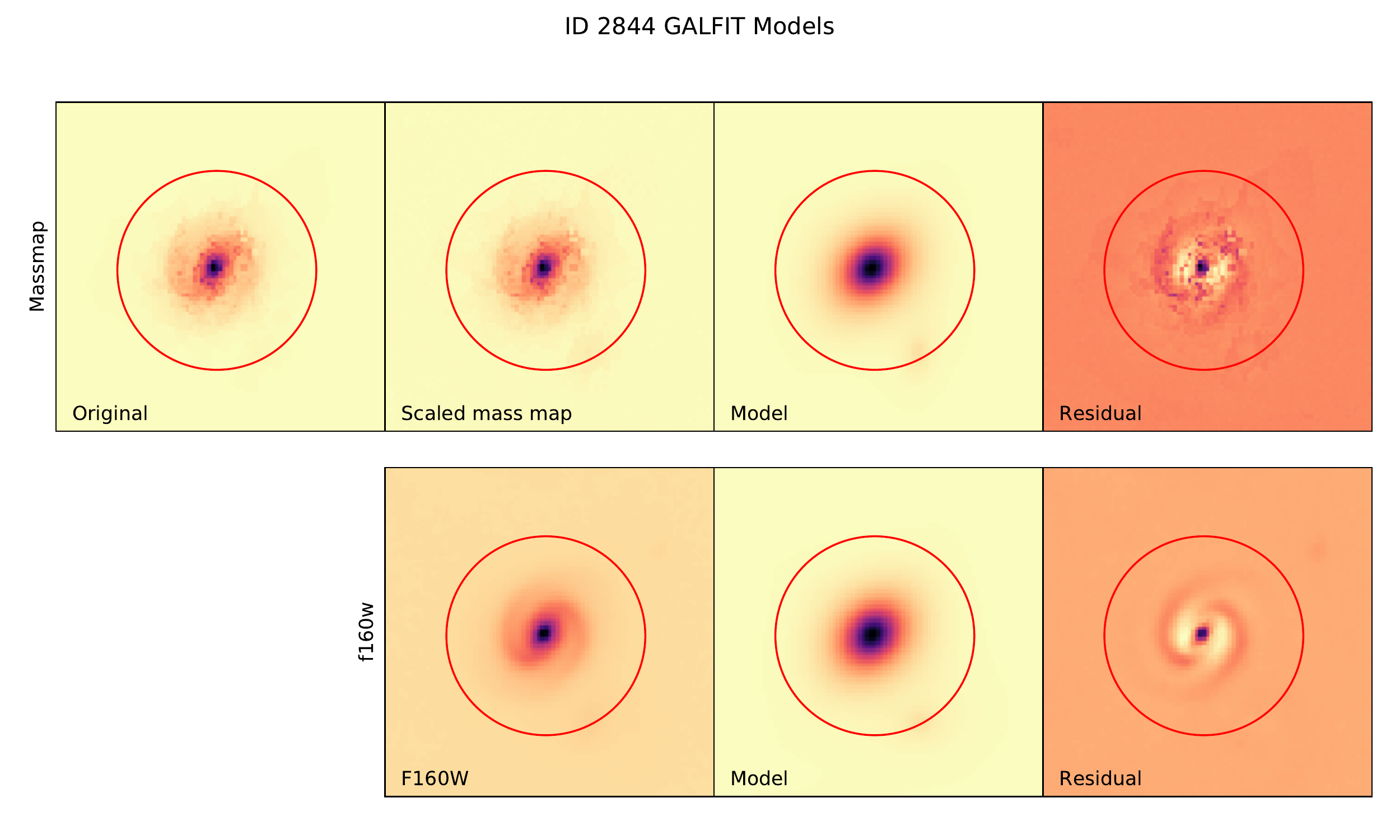}
	\end{subfigure}
	\begin{subfigure}
	\centering
		\includegraphics[width=0.75\textwidth]{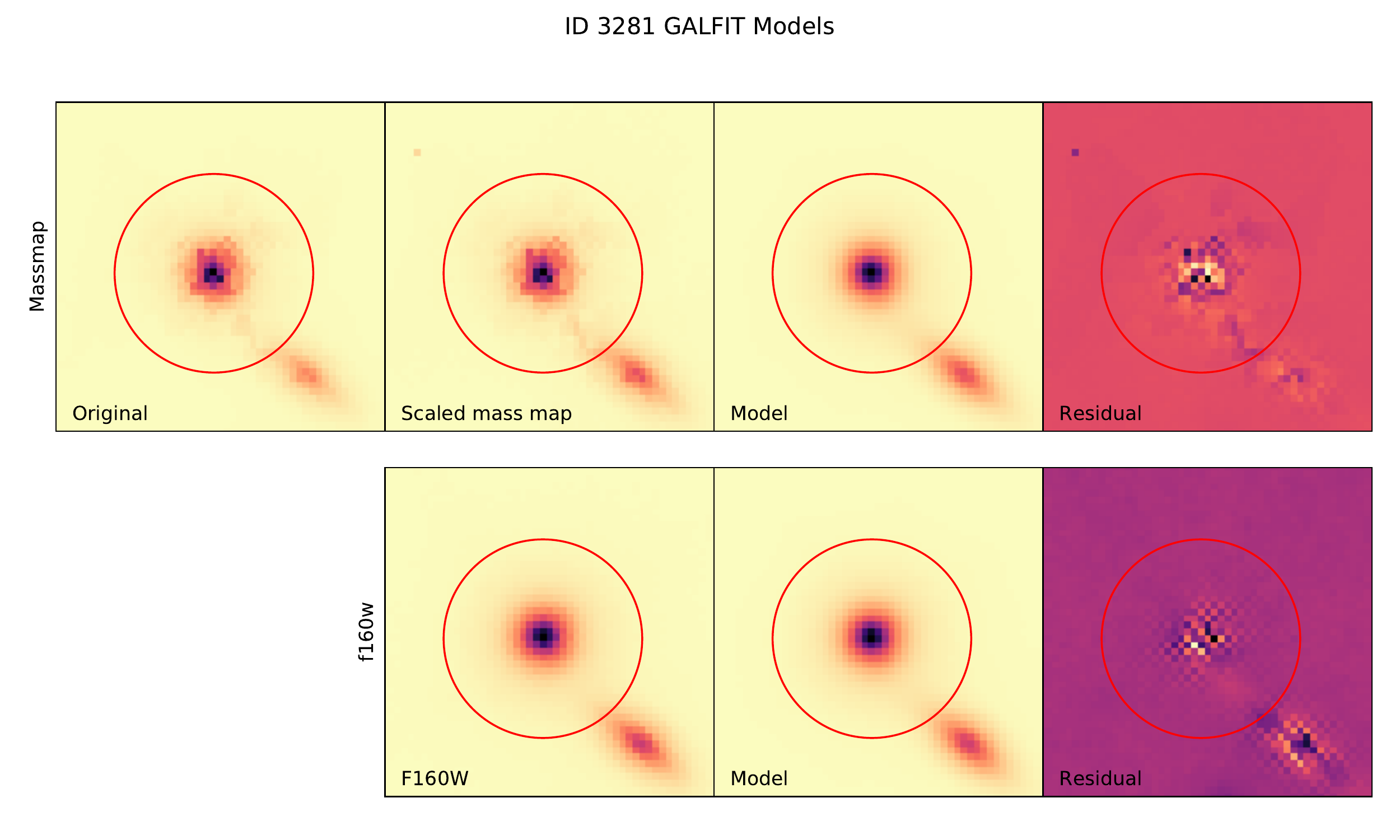}
	\end{subfigure}
	\caption{Top: An example fit from a field galaxy. Bottom: Example fit from a cluster galaxy. The circle denotes 3$\times$ the effective radius as defined by SExtractor. Note that structures such as spiral arms are not part of the S\'ersic fit and end up in the residual if functions specifically made for describing spiral arms are not fit. Bin residuals of stellar mass maps also end up in the residual image.} \label{fig:galfit_fig}
\end{figure*}
\begin{sidewaystable}
\centering

\begin{tabular}{| c | c c c | c c c | c c c | c c c |}
\hline
 \hspace{1pt} & \multicolumn{6}{c|}{\textbf{Quiescent Galaxies}}  &  \multicolumn{6}{c|}{\textbf{Star-forming} \textbf{Galaxies}}\\
 \hline
 \hline
  & \multicolumn{3}{c|}{$n$} & \multicolumn{3}{c|}{$R_{eff}~[kpc]$} & \multicolumn{3}{c|}{$n$} & \multicolumn{3}{c|}{$R_{eff}~[kpc]$ } \\ 

  \textbf{log(\mstar/\msun)} & \textit{F160W} & \textit{$\Sigma_\star$} & \textit{\% diff} &  \textit{F160W} & \textit{$\Sigma_\star$} & \textit{\% diff} & 
   \textit{F160W} & \textit{$\Sigma_\star$} & \textit{\% diff} &  \textit{F160W} & \textit{$\Sigma_\star$} & \textit{\% diff}  \\
   \hline
   \hline
  \textbf{8.5-9.0} & $2.10\pm0.12$ & $1.95\pm0.10$ & (-7.27) & 
  $1.49\pm0.08$ & $1.35\pm0.06$ & (-9.19) & 
  $1.27\pm0.09$ & $1.36\pm0.13$ & (6.42) &
  $2.39\pm0.14$ & $2.24\pm0.12$ & (-6.13) \\   

  \textbf{9.0-9.5} & $2.52\pm0.10$ & $2.79\pm0.16$ & (10.46) & 
  $1.57\pm0.12$ & $1.42\pm0.08$ & (-9.65) &
  $1.27\pm0.09$ & $1.32\pm0.08$ & (4.41) & 
  $2.71\pm0.13$ & $2.50\pm0.11$ & (-7.69) \\
 
  \textbf{9.5-10.0} & $3.18\pm0.12$ & $3.39\pm0.18$ & (6.54) & 
   $1.52\pm0.16$ & $2.16\pm0.16$ & (-9.04) &
   $1.52\pm0.15$ & $1.59\pm0.13$ & (4.75) &
   $2.99\pm0.17$ & $3.02\pm0.30$ & (1.23) \\
   
  \textbf{10.0-10.5} & $3.48\pm0.12$ & $3.61\pm0.18$ & (3.72) & 
  $1.86\pm0.20$ & $2.54\pm0.16$ & (-16.87) &
  $1.86\pm0.32$ & $1.97\pm0.27$ & (5.97) & 
  $3.42\pm0.33$ & $3.65\pm0.70$ & (6.75) \\
  
  \textbf{10.5-11.0} & $3.58\pm0.16$ & $3.80\pm0.25$ & (6.14) &
  $3.41\pm0.36$ & $2.66\pm0.22$ & (-22.14) &
  $1.01$        & $0.86$        & (-15.53) & 
  $2.02\pm0.03$ & $1.65\pm0.05$ & (-18.54) \\
  
  \textbf{11.0-11.5} & $3.97\pm0.39$ & $3.86\pm0.55$ & (-2.77) & 
  $3.60\pm0.76$ & $3.57\pm0.65$ & (-0.62) &
  \nodata & \nodata & \nodata &
  \nodata & \nodata & \nodata \\
  
  \textbf{11.5-12.0} & $3.92\pm0.68$ & $3.92\pm1.22$ & (0.09) &
  $2.65\pm0.32$ & $3.04\pm1.05$ & (14.56) &
    \nodata & \nodata & \nodata &
  \nodata & \nodata & \nodata \\
  
   \hline
\end{tabular} 
\caption{Table displaying values from Figure \ref{fig:IRvsSM2}. There are bin-averaged parameters along with their standard error, and the percent difference between $H-$band light profile and $\Sigma_\star$ distribution profile is also shown.\label{table:IRvsSM2side}} 
\end{sidewaystable}

\subsection{A S\'ersic model of a mass profile}\label{sec:method-GALFIT}
The S\'ersic index and effective radius for each galaxy's F160W flux density and stellar mass distribution is obtained using GALFIT \citep{P2002,Peng:2010}.  GALFIT provides parametric fits to galaxy light profiles provided that background estimation is done carefully. The postage stamp cutouts of the selected sample in every HST band were made with the dimensions of the cutout being \textit{20 times} the half-light radius of the galaxy as defined by SExtractor, so as much of the galaxy's light, or in this case, stellar mass, can be captured and compared to the background ``flux". 

 For each object's F160W cutout image and its stellar mass distribution, a single S\'ersic profile with varying $n$ is fitted. The central object in each cutout in Figure \ref{fig:galfit_fig} is a galaxy that is part of our sample. Since GALFIT was created for fitting surface brightness profiles, many of the parameters it fits are related to light profiles, such as magnitude zeropoint. This makes it relatively simple to fit a single component S\'ersic profile to each F160W cutout, the only point of failure for the surface brightness case being from not including the right amount of ``secondary" objects in each cutout. If too many secondary objects are fit, the solution will not converge, and conversely, if not enough background objects are fit, the model computed by GALFIT is less accurate due to contamination from those sources. For the stellar mass distributions, scaling is performed to make the image more compatible with GALFIT, even though it is a program designed primarily for light profiles, which will be explained in the following sections.

\subsubsection{Optimizing stellar mass maps for GALFIT}\label{subsec:galfit2}

A magnitude limit was set for the fainter background objects that are not our primary sample to be either $<30$, or $<m_{obj} + 5$, whichever is the brighter limit, where $m_{obj}$ is the magnitude of the central object in the $H$-band (F160W). GALFIT also requires a sigma image to determine the ``brightness" of the background and the extent of the S\'ersic profile for each object. GALFIT is capable of generating a sigma image by taking the sigma at each pixel from both the source and from a uniform sky background, which is then summed in quadrature. The background pixels are defined by the segmentation map created by SExtractor using the detection filters of the Frontier Fields.

The actual stellar masses in each pixel of the mass maps are large numbers ($10^3 - 10^7$), typically not used with GALFIT. Therefore, a scaling of ($\Sigma_i$ F160W flux)/($\Sigma_i$ Mass), summed over every pixel in the postage stamp, was applied to the stellar masses before GALFIT was run. This re-normalization does not affect any of the GALFIT output parameters except ``total magnitude". The constant scales the mass maps in such a way that the ``zeropoint" parameter can be used from the F160W flux zeropoint.

Gaussian noise of similar levels to the F160W band's background noise was also added to the stellar mass maps with the mean and standard deviation scaled relative to the overall stellar mass. The added noise does not affect the total stellar mass of the object, as can be seen in Figure \ref{fig:galfit_fig}, but it is a necessary aid for GALFIT to properly find objects to fit in the mass maps. Without the inclusion of the noise, the GALFIT algorithm overestimates the galaxies' intensity; even small spikes in the mass map are highly significant and can cause GALFIT to be unstable. This streamlines the galaxy modelling process, because the same parameter file for the F160W cutouts can be used on the mass map of the same object. 

The factor by which the stellar mass is scaled would also mean the secondary objects will have very similar mass ``magnitudes" to their F160W fluxes, so the secondary objects being fit can have the same initial magnitude. The bins for these fainter objects do not make them as well defined as the central object, however they do not need to have accurate S\'ersic profiles since GALFIT is robust enough to compensate for less accurate initial conditions while keeping the primary object as the most optimal fit. Thus, reusing the F160W parameters for secondary objects is sufficient. We place constraints on the S\'ersic index for every single component fit, whether light profile or mass profile. The limits for $n$ are in the range $0 < n < 10$ for non-bCGs, and $0<n<12$ for bCGs. Other parameters are left free, but their initial guesses are the values of the parameters from the HFF catalog.

As mentioned in \S \ref{sec:method-GALFIT}, the cutout sizes need to be 20 times the SExtractor defined half-light radius for background level estimation. For the majority of cases, the cutout does not contain more than 110 objects, which is the maximum number of objects that GALFIT can fit \citep{vanderWel:2012}. However, not every source needs to be fit with a S\'ersic profile for proper background estimation and convergence, only objects with a magnitude brighter than 30 need to be fit due to the noise levels of the background. The resulting S\'ersic models for both the brightness fit and the mass fit are shown in Figure \ref{fig:galfit_fig}. From the original 400 cluster galaxies, 18 bCG galaxies from Abell 1063 were removed from the sample because they could not be recovered from the original images after their removal from the science images. Out of the remaining 382 cluster galaxies, 368 have successful stellar mass S\'ersic profile fits, and 92 out of the 96 field galaxies have successful stellar mass S\'ersic profile fits. The failures had no output from GALFIT and their parameters failed to converge on a proper fit. The objects that fail are not biased towards any stellar mass, since the percentage of failed fits for $\mstar < 10^{9.5} \msun$ is roughly 9.5\%, and the percentage of failed fits for $\mstar > 10^{9.5} \msun$ is roughly 9.1\%.

\subsection{Surface Brightness versus Stellar Mass Profiles}\label{sec:res1}
\begin{figure*}[t]
	\centerline{\includegraphics[width=0.8\textwidth]{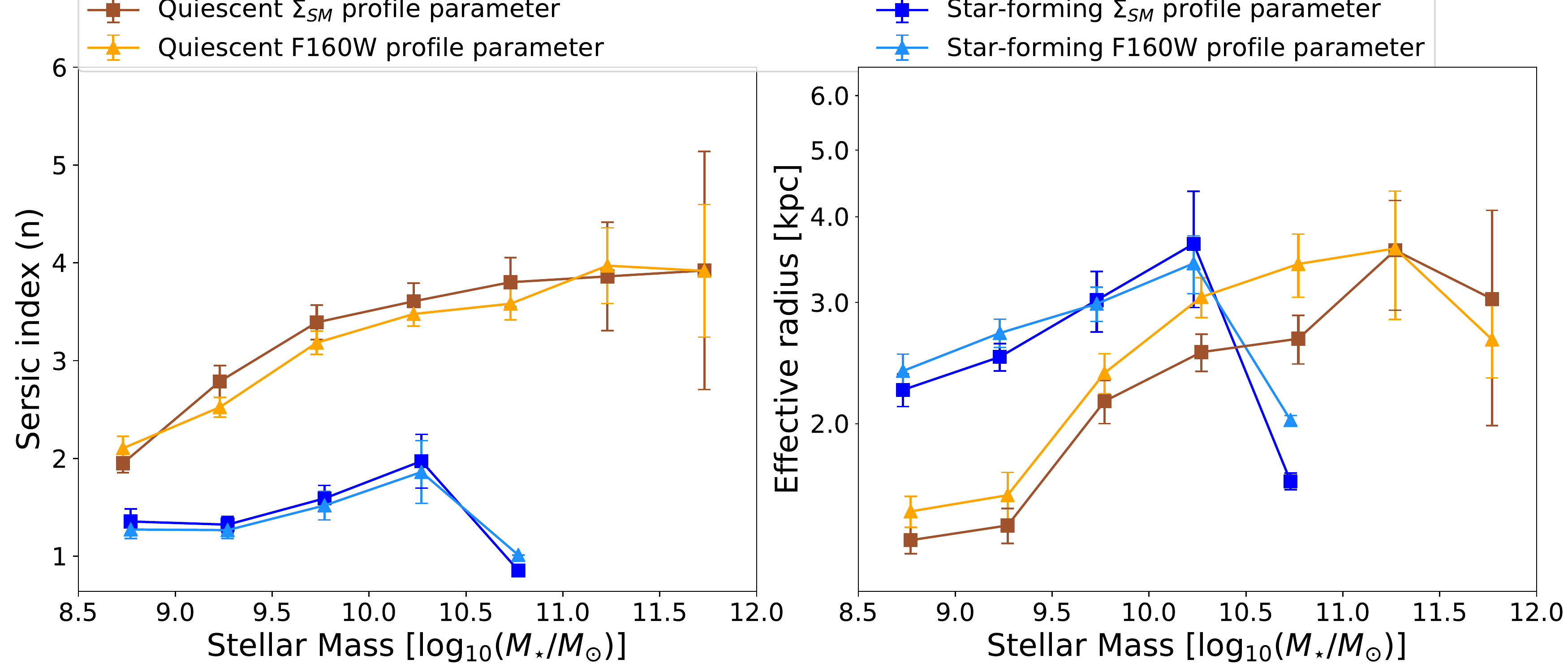}}
	\caption{S\'ersic index vs Stellar mass and Effective radius vs stellar mass is plotted to show how S\'ersic index and size is affected by outshining bias in the infrared. Each point is the average parameter for 0.5dex mass bin. Error bars represent standard error for the points in each mass bin. Star-forming and quiescent sample is defined according to UVJ colours outlined in \S\ref{sec:data-uvj}. S\'ersic indices are generally larger for mass profiles and effective radii are generally larger for F160W light profiles.}\label{fig:IRvsSM2}
\end{figure*}

Infrared bands such as F160W are sometimes used as a proxy for stellar mass. There is a steep outshining bias for more massive stars in the UV, but the outshining is much less in the IR for stars of any mass, making IR photometry a better overall estimate of stellar mass. However, F160W and stellar mass are not necessarily identical, and the full SED of a galaxy is still the most accurate way to obtain its integrated stellar mass (i.e. see \citealt{W2012} for another demonstration of this difference).

Figure \ref{fig:IRvsSM2} compares average S\'ersic index and effective radius between the light profile and the stellar mass profile. Each bin is 0.5dex of stellar mass, with error bars representing standard error. $n_{F160W}$ denotes the average S\'ersic indices derived from F160W brightness profile and $n_{mass}$ for the stellar mass density ($\Sigma_\star$) profile. This is also displayed in Table \ref{table:IRvsSM2side} which displays the numerical values of the average parameters in Figure \ref{fig:IRvsSM2} and their errors. This means the differences in S\'ersic index between mass profiles and F160W brightness profiles are on the whole not statistically significant. For percentage differences, the overall trend for quiescent galaxies is a 2.4$\pm0.8$\% increase in S\'ersic index and a 1.2$\pm1.7$\% increase in S\'ersic index for star-forming galaxies. This means the F160W profile's S\'ersic indices ($n_{F160W}$) can be said to reasonably represent the underlying stellar mass distribution's S\'ersic indices ($n_{mass}$), although they are not analogous. 


In terms of percentage differences between half-light and half-mass radii, the average effective radius for quiescent galaxies is 7.6$\pm1.6$\% lower for the stellar mass maps compared to F160W flux, and 4.9$\pm1.5$\% lower for star-forming galaxies, where the errors are standard errors of the mean. For quiescent galaxies, the average difference between half-mass radii ($R_{mass}$) and half-light radii($R_{F160W}$) is more constant as a function of stellar mass. Whereas for star-forming galaxies, the difference is not as pronounced except for the highest stellar mass bin, with a 22.1\% difference between $R_{F160W}$ and $R_{mass}$ at $10.5<\log(\mstar/\msun) <11$. This seems to indicate the outshining bias for F160W IR light affects the effective radius more than the S\'ersic index. \cite{S2013} found an overall 25\% decrease between the half-light radius and half-mass radius, which is much greater than our overall percentage decrease, but their redshift range is $0.5 < z < 2.5$. Their sample includes galaxies of a much higher redshift than this work. Galaxies at high redshift, especially around $z \sim 2$ are undergoing much higher rates of star formation than at later times, which would mean higher flux but less stellar mass. For galaxies at $0.25< z < 0.6$, less star formation should be taking place overall. This is consistent with the framework where there are lower merger rates at late cosmic times, which leads to both a lower SFR, and also the smaller decrease from $R_{F160W}$ to $R_{mass}$.
\section{Analysis of mass profile parameters}\label{chap:results}

\begin{figure*}
	\centerline{\includegraphics[width=0.8\textwidth]{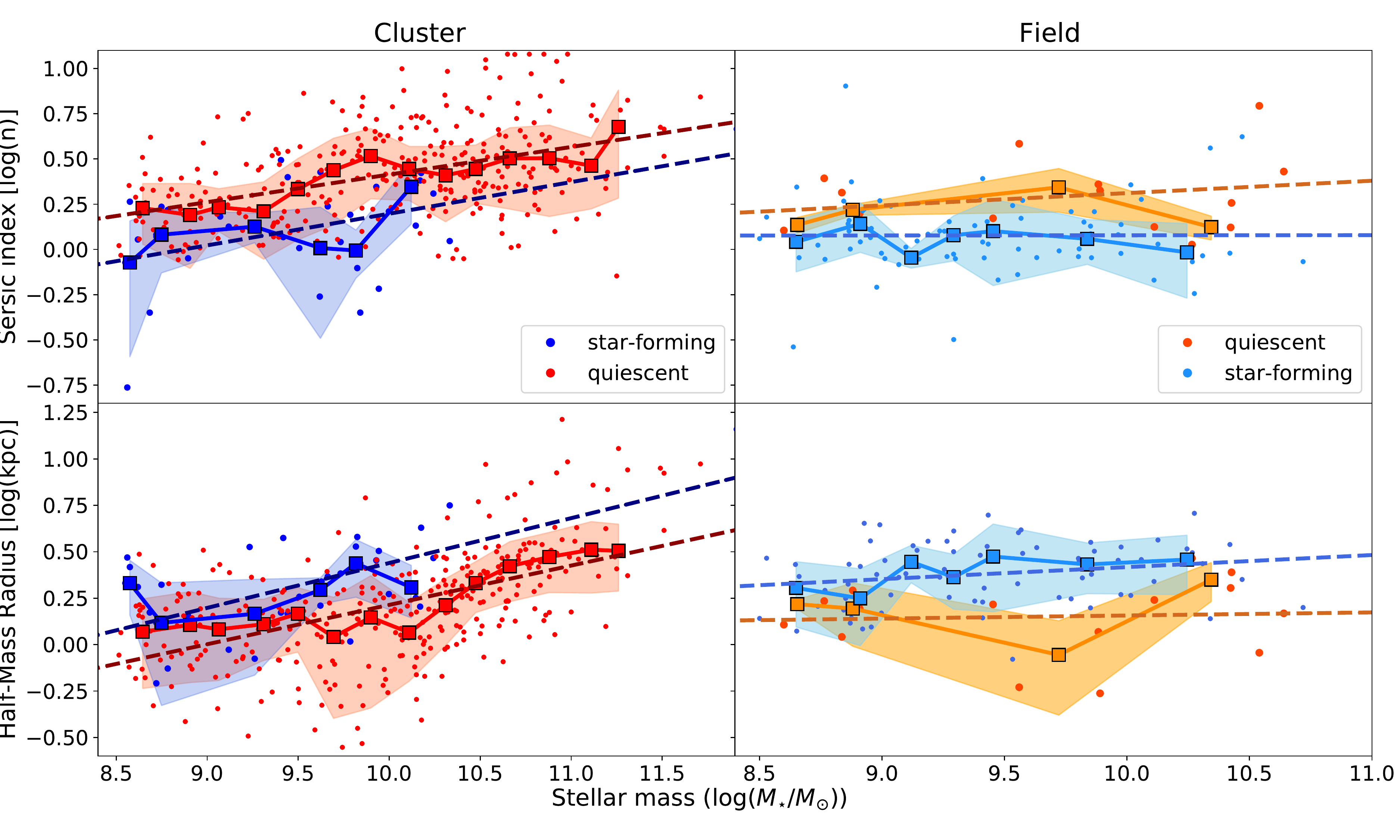}}
	\caption{Log Stellar mass vs Log S\'ersic index (top) and stellar mass vs size (bottom) plots for star-forming and quiescent galaxies in both the cluster and field. The trend lines with square points represent the \textit{median} of groups of five data points for star-forming field (light blue), star-forming cluster (dark blue), and quiescent cluster galaxies (red), and groups of three for quiescent field galaxies (orange). The shaded regions represent one standard deviation from the trend line in each mass bin. The dashed lines are linear lines of best fit to each sample's trend line. \label{fig:allparams}}
\end {figure*}

In this section, we present the analysis of the stellar mass density profile's parameters to classify the morphology of star-forming and quiescent galaxies in the cluster member sample. Through using parameters that describe the stellar mass distribution (which will be denoted $\Sigma_\star$ as it is a surface density), we aim to draw links between certain galaxies that have different star-formation rates, and live in different environments, but have similar morphologies. We use two S\'ersic parameters of mass profiles fits, specifically S\'ersic index ($n_{mass}$) and half-mass radius ($R_{mass}$), to quantify the morphological differences in galaxies. The galaxy sample of 368 cluster members and 92 field galaxies is divided into four groups based on their star-formation activity, and whether they live in a cluster (dense) or field (less dense) environment. 
  

\subsection{Mass profile differences in quiescent cluster galaxies}\label{sec:res2}

\fontsize{8}{8}\selectfont
\begin{table}[H]
\centering
	\caption{\textit{Slopes of the lines of best fit in Figure} \ref{fig:allparams} \label{res:slopetable}}.
	\begin{tabular*}{\columnwidth}{l l l c}
	\hline
	 & & cluster & field \\
	 \hline
	S\'ersic & SF & $4.85\pm1.31 \times10^{-1}$& $-0.63\pm1.73\times10^{-1}$\\
	index & Q & $1.13\pm0.41$ & $5.12\pm4.21\times10^{-1}$ \\
	\hline
    half-mass &	SF & $2.26\pm2.68\times10^{-7}$ & $1.14\pm2.54\times10^{-3}$ \\
    radius & Q & $2.57\pm4.20\times10^{-9}$ & $2.43\pm3.28\times10^{-2}$ \\
	\hline
	\end{tabular*}
\end{table}
\normalsize

In Figure \ref{fig:allparams}, the S\'ersic index $n_{mass}$ and half-mass radius $R_{mass}$ of each galaxy are plotted against their stellar mass and assigned an environment-star-formation classification. The plots in Figure \ref{fig:allparams} demonstrate how the S\'ersic parameters depend on stellar mass, environment, and star-formation activity. The distribution of points for quiescent galaxies overlap with the distribution of star-forming galaxies, for both the cluster and the field. The shaded regions are $1\sigma$ from the box-kernel generated trend line average, and there are star-forming and quiescent data points which lie in both shaded regions.

The panels in Figure \ref{fig:allparams} also have lines of best fit for each mass-S\'ersic index and mass-size relation shown. For star-forming galaxies in the field (right panel), the S\'ersic index $n_{mass}$ line of best fit has a slope close to zero. This shows the SF Field $n_{mass}$ is independent of stellar mass. The same is not true for cluster star-forming galaxies (left panel), for which line of best fit's slope has a value of 0.485$\pm0.131$ and shows a slightly increasing relation with stellar mass. Both quiescent galaxies in the cluster and the field have $n_{mass}$ increasing with stellar mass, but the slope for quiescent cluster galaxies is 1.13$\pm0.41$ (see Table \ref{res:slopetable} for the slope of all the lines of best fits in Figure \ref{fig:allparams}). Although the median points and their error bars (the shaded regions in the top left panel of Figure \ref{fig:allparams}) indicate that there shouldn't be a statistically significant difference between the two slopes. The dispersion of $n_{mass}$ for the quiescent cluster sample is much larger than the quiescent field sample, with the distribution becoming wider with increasing stellar mass, which could explain the larger slope in the line of best fit. The large distribution of S\'ersic indices for quiescent cluster galaxies compared to all other populations may indicate that these galaxies have undergone diverse pathways of quenching. While there are plenty of quiescent cluster S\'ersic indices greater than 4, meaning they are likely to be elliptical galaxies, there is \textit{also} a sizable number of these galaxies which have $n_{mass} \lesssim 2$. Since a higher S\'ersic index indicates a more bulge-dominated morphology, this has implications for how stellar mass affects morphology in dense cluster environments. \S\ref{subsec:sef-groups-2} explores the morphology of quiescent cluster galaxies further.

Half-mass radius ($R_{mass}$) versus stellar mass is also shown in the bottom row of Fig.~\ref{fig:allparams}. The $R_{mass}$ values are the GALFIT derived $R_{\text{eff}}$ for the stellar mass maps,  which is the half-mass radius along the semimajor axis. Unsurprisingly, for each classification, $R_{mass}$ increases with stellar mass, but at different rates. The trend lines show that on average, star-forming galaxies are larger than quiescent galaxies. This is consistent with previous results which used half-light radii \citep{V2014,Mowla:2019,Nedkova:2021} However, looking at the individual points, there is also considerable overlap between star-forming and quiescent galaxies just as with the S\'ersic index distributions. This is shown in the shaded regions, indicating 1$\sigma$ from the trend line. One of the most interesting features of the mass-size relations is that the cluster star-forming and quiescent sizes have a large overlap in distribution. Looking for quiescent cluster galaxies which have small $n_{mass}$ but a large $R_{mass}$ could point to a population of environmentally quenched galaxies that retained a disk morphology. 



\subsection{S\'ersic parameters that change with both mass and environment}\label{subsec:sef-groups-2}

\begin{figure*} 
	\centerline{\includegraphics[width=0.8\textwidth]{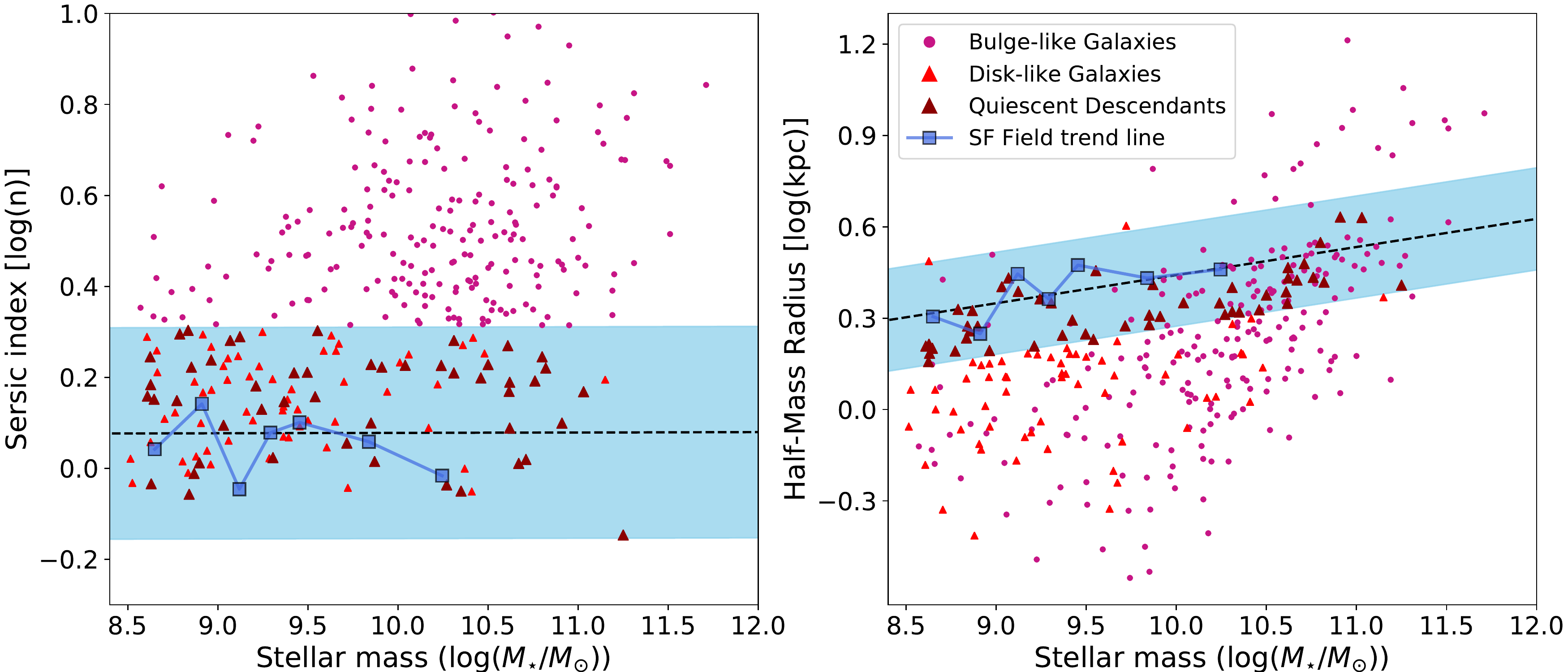}}
    	\caption{\textit{Mass-S\'ersic index} and \textit{Mass-Size} relation plots for quiescent cluster galaxies. The small red triangles are the quiescent cluster galaxies with ``disk-like" S\'ersic indices. They are within one standard deviation of the star-forming field population's line of best fit, shown as the dashed black line, with the shaded blue region around it indicating the 1$\sigma$ region. The large red triangles are quiescent cluster galaxies that are \textit{morphologically consistent} with star-forming disk galaxies in the field -- they have ``disk-like" S\'ersic indices \textit{and} ``disk-like" half-mass radii. The half-mass radii selection was also defined as within one standard deviation (1$\sigma$) of the average half-mass radius line of best fit for the star-forming field population (dashed black line on the Mass-Size relation plot). Galaxies that are not disk-like in S\'ersic index are represented by purple points and are designated ``bulge-like".}  \label{fig:focusgalaxies}
\end {figure*} 
\begin{figure*} 
	\centerline{\includegraphics[width=0.8\textwidth]{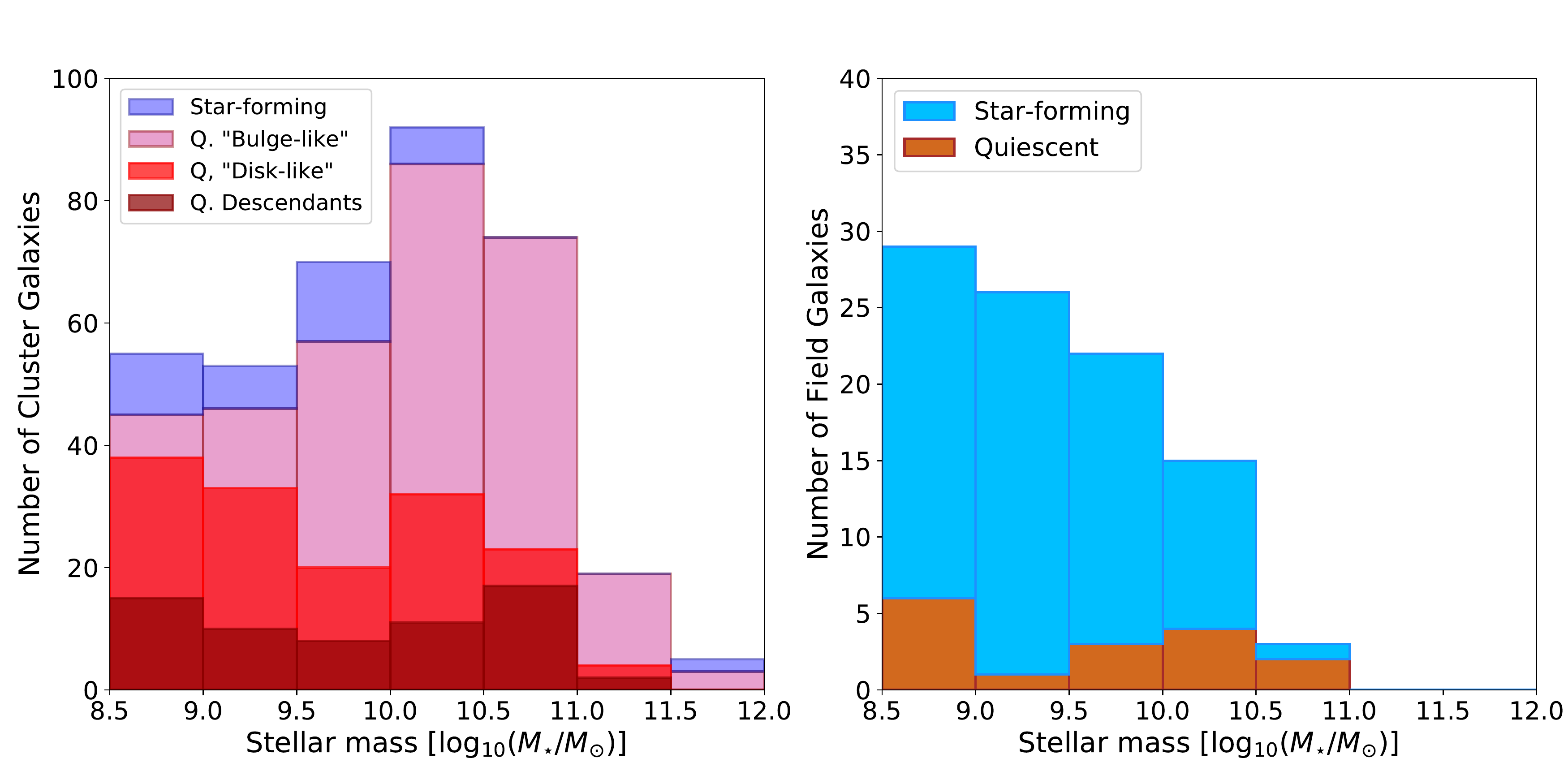}}
	\caption{Histogram breakdowns according to star-formation rate for cluster (left panel) and field (right panel) populations. The cluster population's quiescent fraction is also separated by morphology into bulge-like, disk-like, and morphological analogues of SF field galaxies. This is based on the same selection criteria as Figure \ref{fig:focusgalaxies}.} \label{fig:selectedhist}
\end{figure*}

The intent of this study is to use statistical tools to look for a population of quiescent cluster galaxies that may be descendants of SF field galaxies that fell into the cluster. Therefore, the next step is to break down the distribution by stellar mass bins and then examine each bin's S\'ersic parameters. This would be evidence that we can use to draw a direct evolutionary link between populations of galaxies in the field to populations of galaxies in the cluster.
The main purpose of generating the mass profiles is to separate the quenched cluster galaxies into distinct populations based on morphology. Figure \ref{fig:allparams} demonstrated that the spread in the distribution of S\'ersic parameters can overlap between quiescent cluster galaxies and star-forming field galaxies. The vast majority of bulge-like quiescents appear to be massive and smaller in size, although a number of disk-like galaxies are also smaller than the field average as well. If quiescent cluster galaxies that are disk-like skew towards a certain mass limit, it would indicate that environmental quenching still depends on stellar mass.

\subsubsection{Quiescent galaxies with disk-like morphology}\label{subsec:params}
If our assumption that environmental quenching does not alter the stellar mass distribution of a galaxy holds, then there must be a subset of quiescent cluster galaxies with \textit{both} their $n_{mass}$ and $R_{mass}$ similar to those observed in star-forming galaxies, the so-called ``disk-like" quiescent galaxies. As we can see in Figure \ref{fig:allparams}, quiescent cluster galaxies have a wide range of S\'ersic index values from 0.5 to 12, and their range of half-mass radii is wider than star-forming cluster galaxies. 

We choose quiescent cluster galaxies which are within 1$\sigma$ of the star-forming field sample's line of best fit to be the sample of ``disk-like" quiescent galaxies. This is shown in Figure \ref{fig:focusgalaxies}. Since the best fit line is almost constant, the 1$\sigma$ can be taken to be the standard deviation of the field galaxies' $n_{mass}$ distribution, as it does not depend on stellar mass.

Within that disk-like quiescent sample, there are a subset of galaxies which are within 1$\sigma$ of the best fit line for the star-forming field's mass-size relation as well. The 1$\sigma$ value is the standard deviations of the star-forming field galaxies' S\'ersic parameters. These galaxies, as shown by the dark red triangles in Figure \ref{fig:focusgalaxies}, fit within the statistics of the SF Field's morphological parameters. This means they are best candidates for having undergone environmental quenching because they can reliably be traced as having originated in the field (sharing morphological features with the field star-forming population) but they show up in the cluster with quenched colours. These galaxies will be referred to as ``\textit{Morphological Analogues of Star-forming Galaxies}". 
 
Figure \ref{fig:selectedhist} more clearly shows the mass dichotomy between the disk-like and bulge-like quiescent galaxies by plotting a stacked histogram of the fractions that both types occupy in their stellar mass bin. Disk-like quiescents make up the majority of the quenched fraction from $10^{8.5} \msun$ to $10^{9.5} \msun$, while bulge-like quiescents dominate at higher masses, from $10^{9.5} \msun$ to $10^{11} \msun$. In Figure \ref{fig:selectedhist}, the disk-like quiescent galaxies when plotted against the entire quiescent cluster distribution appear in much greater numbers at the low mass end. Altogether the disk-like quiescent galaxies make up roughly 42\% of the quiescent cluster sample, with $\sim 16\%$ being selected as the quiescent morphological analogues to star-forming galaxies. The majority of the morphological analogues to star-forming galaxies occur on the very low mass end of the quiescent cluster population, below $10^9 \msun$. The smallest and largest mass bins notably only contain quiescent galaxies that fall into one classification, with the smallest bin having almost no quiescent galaxies of bulge-like morphology, and likewise the largest bin having no quiescent galaxies of disk-like morphology. 

\begin{figure*}[t]
	\centerline{\includegraphics[width=0.8\textwidth]{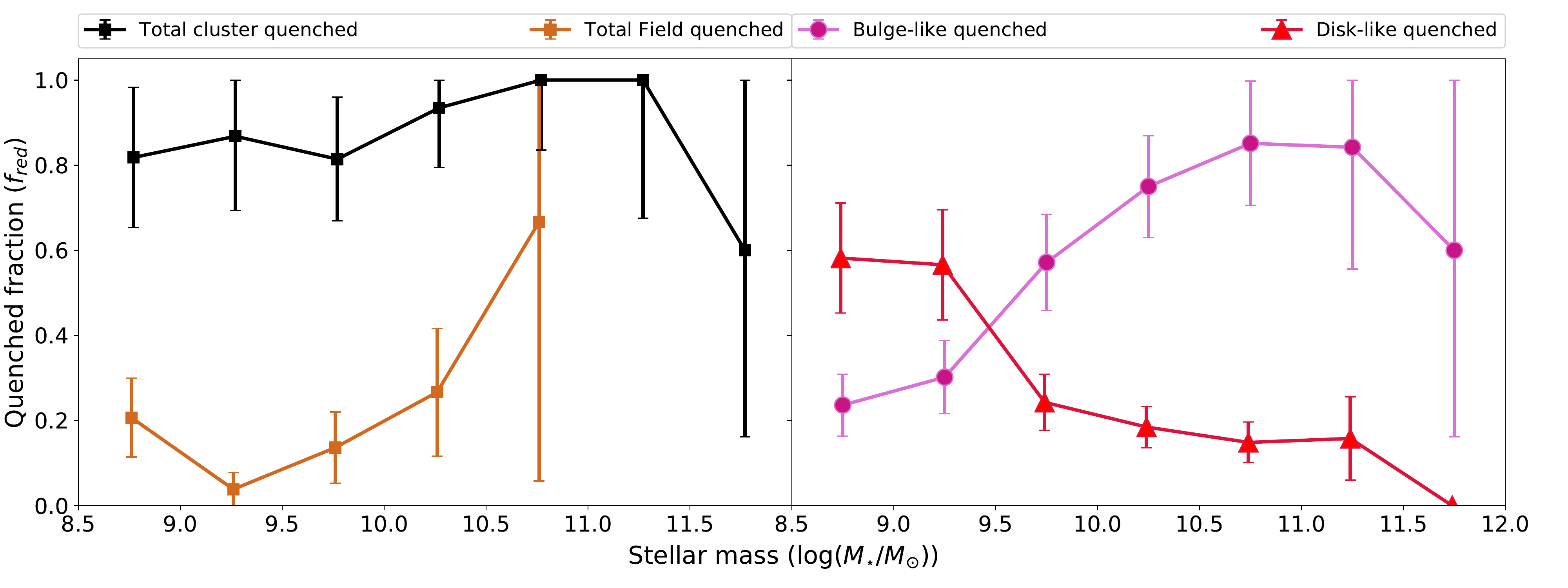}}
	\caption{Left panel: Quiescent fractions for cluster and field.  Right panel: Quiescent fraction for cluster galaxies decomposed according to their morphology into disk-like quenched fraction and bulge-like quenched fraction. Notice the bulge-like quenched fraction for the cluster, and the disk-like quenched fraction for the cluster have opposite trends with stellar mass. Bulge-like $f_{red}$ increases as stellar mass increases while disk-like $f_{red}$ decreases with stellar mass, indicating morphology of quenched galaxies are dependent on $\mstar$. The errors are Poisson errors.}\label{fig:qfrac}
\end{figure*}
\begin{figure*}[t]
	\centerline{\includegraphics[width=\textwidth]{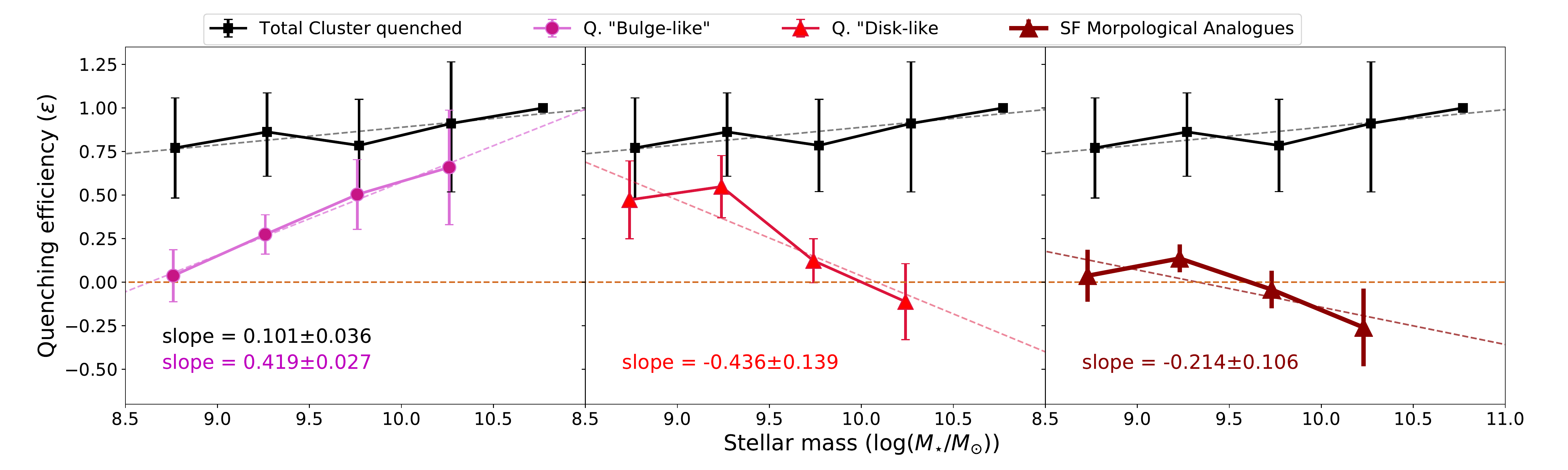}}
	\caption{Environmental quenching efficiency vs. stellar mass plots for three quiescent populations in the cluster: bulge-like, disk-like, and star-forming morphological analogues. The quenching efficiencies($\epsilon_\rho$) are calculated relative to the quenching efficiency of the field. Therefore, the orange dashed line denotes the ``baseline" field quenching efficiency against which the various cluster populations' $\epsilon_\rho$ are plotted. The bulge-like and disk-like quenching efficiencies have similar trends as in Figure \ref{fig:qfrac}. Bulge-like galaxies are more efficiently quenched in clusters at higher masses, while disk-like galaxies are more efficiently quenched in clusters at lower masses. The population of morphological analogues to star-forming galaxies in the rightmost panel follow a similar relationship to disk-like galaxies, being more efficiently quenched at lower stellar masses.}\label{fig:qeff}
\end{figure*}
\subsubsection{Disk-like vs bulge-like quenching efficiencies}\label{subsec:disk-bulge-qeff}


While Figure \ref{fig:selectedhist} shows the numerical breakdown of types of galaxies in each mass bin, the left panel of Figure \ref{fig:qfrac} displays the quenched fraction (denoted $f_{red}$) as a function of stellar mass for the cluster and field samples (black and orange trends respectively). We will refer to quenched fractions as only the points and not the range of their error bars. For the cluster, the total quenched fraction, as well as the quenched fractions for galaxies with bulge-like (magenta symbols) and disk-like (red symbols) morphologies are shown in the right panel of Figure \ref{fig:qfrac}. For the field, only the total quenched fraction is plotted. The total cluster quenched fraction increases with stellar mass until $\mstar \sim 10^{11} \msun$ where the quenched fraction reaches 100\%. The only exception is the last mass bin as there was one massive star-forming galaxy with a mass $> 10^{11.5} \msun$.

The field $f_{red}$ stays below 30\% until its largest mass bin of $10^{10.5} \msun$, unlike the cluster $f_{red}$, which is above 80\% save for the smallest mass bin ($<10^{8.5} \msun$). However, at $10^{10.5} \msun$, the field  $f_{red}$ reaches 100\%, which is one mass bin lower than the cluster $f_{red}$. When looking at the quenched fraction for cluster galaxies, there are unexpected results. If environmental quenching were not mass-dependent, then we  would expect both bulge-like and disk-like quenched fractions to have the same relation with stellar mass and \textit{not} opposing trends. 

This result can be compared with the star-forming fractions in \cite{Varma:2022}, where they examine the morphologies of galaxies from redshift 3 to 0.5, using an observational sample of galaxies from CANDELS, and a simulated sample from TNG50 \citep{Nelson:2019,Pillepich:2019}. They find that at a redshift of $0.3 < z < 0.7$ for observations, and $z=0.5$ for simulations, that for disks and disks with spheroids, the star-forming fraction increases with increasing stellar mass for both simulations and observations. For pure bulges, it stays roughly constant with stellar mass. \citeauthor{Varma:2022}'s sample of galaxies mostly reside in lower environmental densities, and yet our quenched fractions for the high-density cluster sample seem to also follow similar trends with respect to morphology and stellar mass. 

While the fraction of quenched galaxies increases with stellar mass for cluster members with bulge-like morphologies, the opposite trend is observed when considering those with disk-like morphologies (Figure \ref{fig:qfrac}, right panel). The median trend line for quiescent cluster S\'ersic indices in Fig. \ref{fig:allparams} is increasing with stellar mass, and the best fit slope is also positive so this result makes sense. Higher S\'ersic indices become more common at higher stellar masses. 

\subsubsection{Environmental quenching and mass dependence}\label{subsec:env-quenching-mass-dep}

From the results presented in the previous section, it is evident that there is a morphological dichotomy between low mass cluster quiescents, which are more disk-like, and high-mass cluster quiescents, which are bulge-like. To find the underlying cause of this difference, a useful measurement is how efficient the two main avenues of quenching are at transforming galaxies from star-forming to quiescent. \cite{P2010} utilized relative quenching efficiencies to compare mass and environmental quenching for galaxies in zCOSMOS up to $z \sim 1$. They showed that mass and environment have completely independent, or ``separable" effects on quenching populations of galaxies. They arrived at this result by showing that the mass-quenching efficiency (denoted by $\epsilon_{m}$) remains constant regardless of local environmental density, and that environmental-quenching efficiency (denoted by $\epsilon_{\rho}$) also remains constant with stellar mass. The \textit{environmental} quenching efficiency is defined in \cite{P2010} in their Equation 3, as a measure of the difference in the number of quiescent galaxies between two regions of space with different overdensities. We write the equation for environmental-quenching efficiency as
\begin{linenomath*}
\begin{equation}
\epsilon_{\rho}(\rho, \rho_0, m) = \frac{f_q(\rho, m) - f_{red}(\rho_0, m)}{f_{blue}(\rho_0, m)} \label{eqn:qenv_eff}
\;\;,
\end{equation}
\end{linenomath*}
 where $\rho_0$ is the reference density (which is the field density), $f_{red}(\rho_0, m)$ is the red sequence/quiescent fraction at the reference density at a certain stellar mass $m$, $f_{blue}(\rho_0, m)$ is the blue cloud/star-forming fraction at the reference density, $f_q(\rho,m)$  is the quiescent fraction at environmental density $\rho$. 
 
 
 Environmental-quenching efficiency ($\epsilon_{\rho}$) as defined in Equation \ref{eqn:qenv_eff} is plotted as a function of stellar mass $\mstar$ in Figure \ref{fig:qeff}, to show how for fixed environments the fraction and the efficiencies of quenching vary with stellar mass. If $\epsilon_{\rho}(\mstar)$ stays constant, then environmental quenching for those galaxies does not depend on stellar mass. That would mean mass quenching and environmental quenching are two effects that are completely independent and separable. For the cluster, the quenching efficiency is calculated both for the disk-like and bulge-like quiescent cluster populations. Since this sample has only two environments, $\epsilon_{\rho_0}$ is chosen to be the field environment, which is why $\epsilon_{env,field}$ is zero for all $\mstar$.

Although \cite{P2010} probed a similar range of redshifts to this work, the stellar mass range that \citeauthor{P2010}'s sample had at that redshift was only from $10^{10.2} \msun$ to $10^{11} \msun$, while there is a much larger range in this work. In the range of $10^{9.5} \msun$ to $10^{10.5} \msun$, the $\epsilon_{\rho}$ and $f_{q}$ are more or less constant (see also \citealt{Nedkova:2021,Cutler:2022}), while for the range $10^{10.5} \msun$ to $10^{11.5} \msun$, the efficiency $\epsilon_\rho$ is actually quite similar to the relation found in \citeauthor{P2010}'s work. Looking at the $\epsilon_{\rho}$ plots in Figure \ref{fig:qeff}, the quenching efficiency of the entire cluster sample does appear to be relatively constant in the range $10^{10} \msun < \mstar < 10^{11} \msun$. However, this is solely based on one point so the trend is not clear. But the rest of the $\epsilon_{\rho}$ for the total cluster sample contradicts the idea that effects of mass and environment on quenching are entirely separable because the quenching efficiency is decreasing for the entire sample disk-like sample, and the non-zero slope of the quenching efficiencies for the disk-like and bulge-like quiescent cluster galaxies.
 
Recently, papers such as \cite{B2016}, \cite{Li2017}, and \cite{P2019} all seem to suggest that there is a mass-dependent component to environmental quenching, especially as redshift increases. This could indicate that disk-like quiescent galaxies that retain the morphology of star-forming galaxies were environmentally quenched through ram pressure stripping (see \S \ref{sec:disc-RPS}).

Another interesting feature in Figure \ref{fig:qeff} of the $\epsilon_{\rho}-\mstar$ plot is the stellar masses at which quenching efficiencies match the field $\epsilon_{\rho}$ for the two types of bulge-like and disk-like quiescents. For bulge-like quiescent galaxies, $\epsilon_{\rho}$ is lowest when it is at the field efficiency level, and likewise for the disk-like $\epsilon_{\rho}$. The quenching efficiency is either greater than the baseline field $\epsilon_{\rho}$ or they are roughly the same as the field efficiency. If the disk morphology quiescent galaxies were quenched environmentally, and the bulge-like quiescents are quenched by mass, i.e. self quenching, then this means existing in a cluster environment enhances self-quenching as well as environmental quenching. This confirms that the two processes operate at different stellar masses. In Figure \ref{fig:qeff}, the bulge-like quiescents increase in efficiency in the leftmost panel around $\mstar {\sim} 10^{9.5}\msun$. In the middle and right panels, the disk-like, and morphological disks quenching efficiency is enhanced at $\mstar < 10^{9.5}\msun$ while it is at the field level for higher masses.

\section{Discussion}\label{chap:disc}

\subsection{Dichotomy of quenching in cluster environments}\label{sec:disc-cluquench}
We have shown that a sizable portion (around 42\%) of our sample are quiescent cluster galaxies that have stellar mass distributions that fall within 1$\sigma$ of the star-forming field galaxies within their respective mass bins. These disk-like quiescent galaxies dominate the number of galaxies at $M_\star \lesssim 10^{9.5} M_\odot$. However, the majority (around 58\%) of the quiescent galaxies are still morphologically quite different from star-forming galaxies, and they are predominantly at a higher stellar mass. This morphological difference, which is measured with the S\'ersic index and effective radius, is present in both the infrared light as well as the stellar mass distribution. 

For massive galaxies, it is possible that secular processes or feedback are more common, and they change the distribution of stellar mass as they quench the galaxy. Otherwise, the trends presented in Figures \ref{fig:qfrac} and \ref{fig:qeff} between the bulge-like population's quenched fraction and quenching efficiency would not have the observed relation with $\mstar$. It has been shown that AGN feedback increases with total $\mstar$, as well as increasing bulge-to-total ratio (i.e. \citealt{Pfenniger:1990,Rafferty:2006,L2014,Bruce:2016}) As stellar mass increases, the disk-like quenched fraction becomes the minority in clusters. Assuming that stellar mass remains conserved, this implies that self-quenching will come to dominate over environmental quenching. This is supported by the quiescent cluster galaxies' $M-n_{mass}$ relation becoming constant when $M_\star \gtrsim 10^{9.5}M_\odot$. In addition, the mass-size relation of quiescent cluster galaxies (bottom left panel of \ref{fig:allparams}), has a relatively consistent slope in the same mass range, confirming that the constant $n_{mass}$ observed is not a quirk of the distribution, but quiescent cluster galaxies in this mass range seem to be preferentially self-quenched. 

There is significantly less, but still a non-zero amount of bulge-like quiescent objects with stellar masses from $10^{8.5}M_\odot < M_\star< 10^{9.5}M_\odot$. Quenching processes that increase the bulge-to-disk ratio of galaxies can stem from tidal effects. Tidal stripping, as opposed to ram pressure stripping, does act on the stars of an infalling satellite galaxy in a way that would disrupt the star-forming disk \citep{M1984, CZ2004}.  Tidal heating is a process where tidal forces increase the random motion of stars in a satellite (also called kinematic heat), making its shape more bulge-like. There is also the possibility of several quenching methods occurring at the same time for these galaxies, which results in the variety of S\'ersic indices and spread in half-mass radii seen in the cluster quiescent sample. 

In simulations, tidal effects were shown to produce S0s from galaxies with AGN (Seyfert galaxies), while ram pressure stripping primarily affects the outer regions \citep{BV1990}. If tidal effects trigger AGN, and that leads to AGN feedback, then this is a viable quenching pathway for the quiescent galaxies on the more massive end of our sample \citep{B2010, M2010}. Some of the low mass but high S\'ersic index objects may also be red nuggets (low mass compact quiescent galaxies) that formed and quenched at an earlier time and fell into the cluster \citep{Z2015}.

\subsection{Framework of environmental quenching}\label{sec:mass-and-quenching} 
Environmental quenching, mostly through ram-pressure stripping is dominant at low stellar mass, and mass/self-quenching is dominant at high stellar mass. The reason that lower mass galaxies are more likely to be environmentally quenched can be explained by how ram-pressure stripping operates. As \citeauthor{GG1972} explains, the ram pressure ($P_{ram}$) comes from the gas density of the intracluster medium, or ICM ($\rho_{ICM}$), and the velocity of the infalling galaxy (equation 61 in \citeauthor{GG1972}).
\begin{linenomath*}
\begin{equation}
	P_{ram} \approx \rho_{icm} v^2, \label{eqn:ram-pressure}
\end{equation}
\end{linenomath*}
There is a restoring force  from the galaxy's disk that counteracts $P_{ram}$ (equation 62 in \citeauthor{GG1972}), which depends on the surface mass density of both the stars ($\sigma_s$) and the gas ($\sigma_g$).
\begin{linenomath*}
\begin{equation}
	F_{restore} = 2\pi G \sigma_s\sigma_g, \label{eqn:restore-force}
\end{equation}
\end{linenomath*}
The stellar mass maps show that the surface densities of lower-mass galaxies are lower than high-mass galaxies. This seems to explain why the $n_{mass} - \log(\mstar)$ relation (top left panel in Figure \ref{fig:allparams}) for the quiescent cluster population is not just linear, but has two plateaus, with the ``cliff" between the plateaus at $\mstar {\sim} 10^{9.5} \msun$. Since the $R_{mass}-\log(M_\star)$ relationship only has a shallow upward slope, that implies the surface density of quiescent cluster galaxies that have $M_\star \gtrsim 10^{9.5} M_\odot$ is greater, and thus they are less prone to being ram pressure stripped. But the quiescent cluster population is already quenched, so their S\'ersic parameters do not reveal what their precursor properties would have been before quenching occurred. 

Examining the star-forming galaxies both in the field and in the clusters, the star-forming field sample's $n_{mass}$ does not depend on stellar mass while the star-forming field sample's $R_{mass}$ is proportional to $\log{M_\star}$. The star-forming galaxies in the cluster seem to have similar relationships with stellar mass, except the more massive star-forming cluster sample have a slightly higher S\'ersic index ($n_{mass} \sim 2$ instead of $n_{mass} \sim 1.5$). If we assume the precursors to the quiescent cluster sample were morphologically similar to both the star-forming populations in the cluster and the field, then their surface mass density would only have a weak dependence on stellar mass. This implies the possibility that RPS operates equally among all masses of galaxies but other quenching methods dominate over RPS over a certain stellar mass limit. 


\subsection{Ram pressure stripping and morphology}\label{sec:disc-RPS}
\subsubsection{Velocity, stellar mass density, and gravity}

 As a galaxy falls through the ICM, the stars remain in place while the gas is stripped away by interactions of the gas in the galaxy's halo with the gas in the ICM. There is some geometry involved to determine how much of the gas is stripped off the galaxy; edge-on infalling disks would lose $\sim50\%$ less gas than face on disks, according to \cite{A1999}, who applied the model of ram pressure stripping (RPS) described in \citeauthor{GG1972} to N-body simulations of disk galaxies. They find that at a certain radius, if the ram pressure exceeds the restoring force per area, then the gas at that radius will be stripped off. 
 
 At first glance it does not appear that ram pressure stripping should preferentially quench lower mass galaxies since none of the terms in equation \ref{eqn:ram-pressure} or \ref{eqn:restore-force} necessarily depend on mass, only mass densities. However, as demonstrated in \cite{A1999}, the existence of this ``stripping radius" means that lower mass galaxies will be quenched by ram pressure much more easily than higher mass galaxies because the former are smaller, have lower surface mass density, or both. This connects to the results that the disk-like quiescents are mostly lower mass, and being ram pressure stripped would maintain their disk-like morphology.
 
Galaxies of any mass falling into a cluster will have similar velocities, but a less massive galaxy will have a smaller potential well, and the gas can escape it more easily at the same $v$ than from a more massive galaxy with a larger gravitational potential well. Escaping gas would not alter the physical distribution of the stellar mass content within the galaxy, so it keeps the same distribution as it had before it was ram-pressure stripped. The galaxy eventually exhausts any remaining gas left in a short amount of time ( $\lesssim 10^9$ years, i.e. \citealt{We2013,Muzzin:2014,Owers:2019}), and the galaxy is now quiescent. Since $\sigma_s$ remains intact, the stellar mass distribution retains $n_{mass}$ close to 1. Again, the finding that disk-like quiescent galaxies are mostly low mass is supported by the ram pressure stripping model.

\subsubsection{Self-quenching at higher stellar mass}
Our sample of quenched galaxies are dominated by bulge-like morphologies as stellar mass increases. Satellite galaxies of a higher mass would be more likely to self-quench, as ram-pressure stripping would not have much of an effect on getting rid of their cold gas. Feedback from AGN is a likely candidate for mass quenching and star-forming massive galaxies of $M_\star \gtrsim 10^{11} M_\odot$ have been shown to have larger bulge-to-total ratios. This trend holds even for stellar mass surface density distributions \citep{L2014}. 

It could be that most massive galaxies are already quenching or quiescent before they fall into a larger halo (e.g. \citealt{V2008}), but these galaxies should have a mass of $M_\star \gtrsim 10^{11} M_\odot$, which is beyond the upper limit of our maximum stellar mass. Merger rates may not fully explain this, as studies at $z \gtrsim 1$ are unclear whether mergers rates are higher or lower in clusters versus the field \citep{Mc2008,Ma2012,D2017,W2019}. These studies also have different methods of classifying merging pairs. \cite{T2008} finds more potential mergers in group environments than cluster environments, but the study is of relatively recent eras at a redshift of $z\sim 0.37$. \cite{Ma2012} concludes major mergers are not enough to explain the evolution of massive compact quiescent galaxies since $z\sim 2$, although that work only focuses on field galaxies. \cite{Lo2013} finds more minor mergers in cluster environments at $z\sim1.62$ compared to the field, and also argue for galaxy mass assembly in clusters via minor mergers. More research is needed to determine whether higher mass quiescent galaxies in clusters have bulge-like morphologies because of minor mergers.

\cite{We2012} show current quenched fractions do not correlate to current density because the quiescent galaxies would have quenched at an earlier time when the density is different. An important note is that quenched fractions are highly dependent on redshift, as more galaxies become quiescent over time. It is generally accepted that in the local universe, mass and environmental quenching are completely separable, but in the $z\sim1$ universe, these quenching methods are less separable for $M_\star \lesssim 10^{11}M_\odot$ \citep{P2010,M2012,K2017, vanderBurg:2020}. Perhaps the separability of mass and environment on quenching affects lower stellar masses as we approach $z\sim0$. Studying this would require applying our methods to a much larger survey of intermediate and high redshift galaxies.

\subsubsection{Could RPS cause morphological transformation?}\label{subsubsec:RPS-morph}

Given the findings outlined in \S\ref{chap:results}, and the very high overall quenching efficiency, the possibility that there may be morphological transformation at $\mstar > 10^{9.5} \msun$ should be considered. Recent papers have shown there is a link between ram-pressure stripping and increased AGN activity \citep{F2012,Poggianti:2017b}. The GASP survey of local jellyfish galaxies reveal that many of their galaxies host an AGN from studying the emission line ratios and the energy required to ionize the gas in the central regions \citep{Poggianti:2017a, Poggianti:2017b}. However, while they can draw a link between ram pressure and AGN, with evidence that ram pressure can increase black hole accretion rates, there is not much direct evidence that ram pressure causes or induces AGN feedback, which is the mechanism that can lead to bulge-growth and morphological transformation \citep{Rafferty:2006}.

\cite{George:2019} found evidence of a central cavity which had been absent of star-formation for the last $10^8$ years. This could be an effect of ram pressure inducing AGN feedback. So far this has been the most direct evidence of AGN feedback related to RPS. But this is not conclusive evidence that RPS leads to morphological transformation. There is still work to be done in how gas outflows are characterized in galaxies undergoing RPS. Nonetheless, RPS plays a possible role in increasing AGN activity \citep{Marshall:2018, Radovich:2019}.

\cite{Ramos-Martinez:2018} utilized a magnetohydrodynamical simulation which modeled how RPS can produce oblique shocks on a flared disk, causing the ISM of the simulated galaxy to flow towards the central regions. They report this mechanism can lead to bulge growth. Other simulation results include \cite{Ricarte:2020}, where the authors found using RomulusC, a high resolution hydrodynamical simulation, that RPS can trigger black hole accretion in galaxies with $\mstar \gtrsim 10^{9.5} \msun$, and increase the rate of gas accretion as well. But RPS suppresses star formation and black hole accretion before it reaches pericenter (i.e. before the galaxy has reached the center of the potential well). This could lead to AGN feedback, or speed up morphological transformation. However, they note that AGN feedback is not fully understood, and there are various models on how this could work. In their simulation of galaxy clusters, they also mention that most galaxies go through pre-processing in smaller groups before falling into the cluster, which complicates the analysis on quenching and morphological transformation.

From both observations and simulations, a link can be drawn between ram pressure and AGN activity. But due to the puzzling nature of AGN-feedback mechanisms as well as pre-processing, more observations are necessary to confirm whether ram pressure can be said to induce morphological transformation for higher mass galaxies.

\section{Conclusions}\label{chap:conc}

 We created spatially-resolved stellar mass maps of over 400 cluster and field galaxies in the Hubble Frontier Fields through SED-fitting the integrated flux of spatially binned images to investigate the underlying stellar mass. S\'ersic profiles were fit directly to these stellar mass maps in order to quantify their morphology in terms of two parameters: the S\'ersic index ($n_{mass}$), and the effective radius, also called the half-mass radius ($R_{mass}$). This was done to explore the relative roles of self-quenching and environmental quenching under the assumption that environmental processes do not cause significant changes in the underlying stellar mass distribution of galaxies.

\vspace{-2mm}
\begin{itemize}
    \item{The S\'ersic indices obtained from modeling stellar mass profile of galaxies are consistent with those measured from the 2-D $H$-band light profile. The effective radius of stellar mass profiles however are on average smaller by 7.6$\pm1.6$\% for cluster galaxies and 4.9$\pm1.5$\% for field galaxies. There is still an outshining bias in infrared filters, but it is only a small bias. This is more pronounced for higher mass galaxies, which can have up to $22\%$ difference in effective radius between F160W flux and stellar mass.}
	\item Assuming that the quiescent galaxies in the clusters that are morphologically similar to star-forming galaxies in the field are their descendants, we can conclude that at $0.25 < z < 0.6$ there is a mass dichotomy in galaxy quenching, where less massive galaxies are more likely to be environmentally quenched while also retaining a disk-like morphology.
	\vspace{-2mm}
	\item If we take the disk quenching efficiency to be the environmental quenching efficiency, it decreases with increasing stellar mass. Environmental quenching is also dominant at $M_\star < 10^{9.5} M_\odot$, accounting for $\sim75.8\%$ of quenching below that mass limit (see right panel of Figure \ref{fig:qfrac}).
	\vspace{-2mm}
	\item Mass and environmental effects are not separable at $0.25 < z < 0.55$ when it comes to how they correlate with quenching efficiency. In addition, \S \ref{chap:disc} discussed other forms of environmental quenching that can change a disk-like profile into a more bulge-like profile, such as tidal heating. Tidal effects, which are related to the environmental density can also induce self quenching for more massive galaxies, such as AGN feedback.
	\vspace{-2mm}
\end{itemize}
This leads to the conclusion that the effects of stellar mass and environment on quenching are not separable at $0.25 < z < 0.6$. Our results confirm that for environmental quenching, there is indeed a mass dependence, which agrees with the results in \cite{B2016,D2016,Li2017,P2019}. In addition, we also find that there is a morphological dichotomy between low mass and high mass quenched satellites in clusters, which means different methods of quenching dominate at different stellar masses for the regions of the universe with the highest densities. There has been evidence that at redshifts greater than $z \sim 1$, quenching mechanisms are not easily separated into environmental or mass quenching. This work shows that at intermediate redshifts between 0 and 1, these quenching efficiencies are different from the local universe.

\acknowledgments 
This work is based on data and catalog products from HFF-DeepSpace, funded by the National Science Foundation, under Grant Number 1513473, and by Space Telescope Science Institute (operated by the Association of Universities for Research in Astronomy, Inc., under NASA contract NAS5-26555), under grant HST-AR-14302. DM and KVN acknowledge generous support from these two grants.

\bibliographystyle{aasjournal}
\bibliography{article}

\end{document}